# NGTS clusters survey – IV. Search for Dipper stars in the Orion Nebular Cluster

Tyler Moulton,[1][*] Simon T. Hodgkin,[2] Gareth D. Smith,[3] Joshua T. Briegal,[3] Edward Gillen,[4,3][†],
Jack S. Acton,[5] Matthew P. Battley,[5] Matthew R. Burleigh,[6] Sarah L. Casewell,[6] Samuel Gill,[5]
Michael R. Goad,[6] Beth A. Henderson,[6] Alicia Kendall,[6] Gavin Ramsay,[7] Rosanna H. Tilbrook,[6]
Peter J. Wheatley[5]

[1]*Department of Applied Mathematics, Harvard John A Paulson School of Engineering and Applied Sciences, 150 Western Ave, Cambridge MA, 02139, US*
[2]*Institute of Astronomy, Madingley Road, Cambridge, CB3 0HA, UK*
[3]*Astrophysics Group, Cavendish Laboratory, J.J. Thomson Avenue, Cambridge CB3 0HE, UK*
[4]*Astronomy Unit, Queen Mary University of London, Mile End Road, London E1 4NS, UK*
[5]*Department of Physics, University of Warwick, Gibbet Hill Road, Coventry, CV4 7AL, UK*
[6]*School of Physics and Astronomy, University of Leicester, Leicester LE1 7RH, UK*
[7]*Armagh Observatory and Planetarium, College Hill, Armagh, BT61 9DG, N. Ireland, UK*

**ABSTRACT**
The dipper is a novel class of young stellar object associated with large drops in flux on the order of $10 - 50$ per cent lasting for hours to days. Too significant to arise from intrinsic stellar variability, these flux drops are currently attributed to disk warps, accretion streams, and/or transiting circumstellar dust. Dippers have been previously studied in young star forming regions including the Orion Complex. Using Next Generation Transit Survey (NGTS) data, we identified variable stars from their lightcurves. We then applied a machine learning random forest classifier for the identification of new dipper stars in Orion using previous variable classifications as a training set. We discover 120 new dippers, of which 83 are known members of the Complex. We also investigated the occurrence rate of disks in our targets, again using a machine learning approach. We find that all dippers have disks, and most of these are full disks. We use dipper periodicity and model-derived stellar masses to identify the orbital distance to the inner disk edge for dipper objects, confirming that dipper stars exhibit strongly extended sublimation radii, adding weight to arguments that the inner disk edge is further out than predicted by simple models. Finally, we determine a dipper fraction (the fraction of stars with disks which are dippers) for known members of $27.8 \pm 2.9$ per cent. Our findings represent the largest population of dippers identified in a single cluster to date.

## 1 INTRODUCTION

The study of photometric variability in Young Stellar Objects (YSOs) is important for a diverse array of topics: stellar evolution, planet detection, and even planetary formation. A novel class of YSO discovered in the past decade, the dipper, offers a unique opportunity to study the early-time properties of stars and their disks (e.g. Alencar et al. 2010, Morales-Calderón et al. 2011, Cody et al. 2014, Stauffer et al. 2015, Rice et al. 2015, Ansdell et al. 2016, Rodriguez et al. 2017, Ansdell et al. 2018, Hedges et al. 2018, Bredall et al. 2020, Roggero et al. 2021). Dippers are low-mass YSOs that exhibit large drops in flux on the order of 10 to 50 per cent. These dips vary in depth, recur over timescales of hours to days, and can be periodic. The dipper occurence rate is about 20–30 per cent in late K- and

M-type Classical T Tauri stars (CTTS), that is they show evidence of a significant disk, and ongoing accretion. Additionally, they have been classified as Type III Variables (see Herbst et al. 1994, Herbst 2012), with dimming events caused by occulting circumstellar matter. Differences between optical and IR extinction observed during dips suggest these occultations are caused by dusty disk material at, or outside of, the co-rotation radius (Alencar et al. 2010, McGinnis et al. 2015). Bodman et al. (2017) suggest that this material is accreted onto the stars along magnetic field lines (and thus allows dippers to be viewed at a larger range of inclination angles than a disk warp model). The authors further posit that explains why dippers are typically observed as low-mass stars: these objects have sublimation radii shorter than the magnetospheric truncation radius, which means that accreting material contains dust that can occult the star at visible wavelengths.

Bodman et al. (2017) explore disk inclination ($i$, whereby $i = 0$ is face-on, and $i = 90$ is edge-on) in their dipper model, suggesting that magnetospheric truncation of the disk, with a misaligned stellar magnetic field, can warp the disk, and even push material out of the plane into an accretion stream. This would allow the dipping phenomena to occur across a wide range of inclinations, and may help to explain its prevalence. However, Ansdell et al. (2016) identify a dipper with a disk that is viewed face-on and suggest a misalignment between the inner and outer disk regions. Indeed this has been previously observed in two stars (Loomis et al. 2017, Kennedy et al. 2017). Additionally, Pouilly et al. (2021) present time-resolved spectropolarimetry of a dipper, deriving a relatively low inclination of 40-50 degrees for the system.

Surveys for dippers have employed a mixture of detection methods. Roggero et al. (2021) used visual examination of K2 lightcurves. Cody et al. (2014) and Bredall et al. (2020) identified dippers from their location in aperiodicity–flux-asymmetry parameter space, where dippers are quasi-periodic to aperiodic and exhibit high flux-asymmetry. Hedges et al. (2018) employed a machine-learning method to identify dippers based on a training set built from previously labelled objects, leading to a significant increase in the number of known dippers in Upper Scorpius and $\rho$ Ophiucus (using the same K2 dataset as Ansdell et al. 2016 and Bodman et al. 2017). The Orion Complex features a large population of young stars, with active-star forming regions such as Orion A and Orion B hosting a 1-3 Myr population (Da Rio et al. 2010, Megeath et al. 2012) and the oldest populations estimated up to 7-10 Myr (see Kounkel et al. 2018, hereafter K18). Orion makes an ideal region for the study of dippers, given the overlap in ages between its star-forming regions, and those of previously identified dipper populations. Indeed, dippers have been already been identified in the Orion Complex in the mid-infrared with Spitzer (Morales-Calderón et al. 2011), and the near-infrared with UKIRT (Rice et al. 2015).

In this paper, we aim to characterize the dipper population of the Orion Complex. We analyze photometric time-series data from the Next Generation Transit Survey (NGTS), centered on the Orion Nebular Complex. The dataset is discussed in Section 2, along with the construction of our training set from known, classified variables. The identification of new stellar variables is described in Section 3. In Section 4 we describe our Machine Learning approach, measurement of periods, and discuss the identification of stars with disks (with more details given in Appendix A). Finally, in Section 5, we present, and discuss our newly discovered sample of dippers.

## 2 NGTS OBSERVATIONS OF ORION

The Next Generation Transit Survey (NGTS) is a ground-based survey based in Paranal Observatory, Chile (for a detailed review of NGTS, see Wheatley et al. 2018). The survey consists of twelve 20 centimeter f/2.8 telescopes. Each telescope has a 520–890 nm bandpass NGTS filter highly adapted to taking photometry of bright late-K and early-M stars, ideal for identification of dipper objects.

Each field that NGTS observes is $2.8° \times 2.8°$. The field observed on Orion is given approximately by the coordinates $82.4° < \alpha < 85.2°$, $-6.8° < \delta < -4°$. The sources detected by the NGTS pipeline from a stacked and dithered 'master' frame (Wheatley et al. 2018), and limited to an NGTS magnitude of $m_{NGTS} < 16$, are shown in Fig. 1. We can see from Figure 11 in K18 that the NGTS field is centered on the Orion Nebula Cluster (ONC). We refer to the larger structure which all member stars are part of as the Orion Complex.

### 2.1 Lightcurves

The NGTS Orion dataset contains photometric time-series data for 8,957 objects. These data were collected over 213 nights with 123,144 images with 10s exposures taken per object at 13s cadence. As is specified at greater detail in Wheatley et al. 2018, each source is associated with an NGTS ID, coordinates, and time and flux values and errors for the observed object. All NGTS data are reduced via a pipeline operated within the University of Warwick data management system. This pipeline applies the updated version of SysRem (Tamuz et al. 2005) used in WASP (Collier Cameron et al. 2006) to detrend NGTS lightcurves by removing signals that are common to multiple sources.

The lightcurves for these objects were binned in time to 11.5 minutes (i.e. a factor $\sim 53$ decrease from their native 13s cadence) to improve computational efficiency in our analysis. In this step, the binned lightcurve fluxes (and uncertainties) were computed as the inverse variance weighted mean of the individual fluxes (and uncertainties) output by the NGTS pipeline. All subsequent analysis uses these binned lightcurves.

We implemented a stage of data cleaning and vetting to ensure our set of lightcurves comprises only sources bearing sufficient high-quality observations. We first removed 48 sources with fewer than 200 observations. We removed an additional 269 sources with NGTS mean magnitude greater than 16.5 (this corresponds to a median RMS of $\sim 0.05$ mag, see Fig 2), leaving a total of 8640 objects.

### 2.2 Auxilliary data

All NGTS sources were positionally cross-matched with the APOGEE-2 survey of the Orion star forming complex K18, which provides spectroscopic data and derived parameters for 8991 stars, as well as Gaia DR2 astrometry and photometry for 16754 stars. Kounkel et al. (2018) used this dataset to identify distinct groups of YSOs and assigned ages ranging from 1 to 12 Myr in the Orion complex. Of 2056 NGTS sources successfully cross-matched with the K18 catalogue, 1284 are classified as members of an Orion group, 485 are classified as field stars, and 287 are classified as membership unknown.

We also matched the NGTS sources against a diverse collection of all-sky catalogues: Gaia DR2 (Gaia Collaboration et al. 2018), Gaia eDR3 (Gaia Collaboration et al. 2021), the Wide Field Infrared Survey Explorer (WISE, Wright et al. 2010), and 2MASS (Skrutskie et al. 2006). These data are important for our dipper and disk analysis of NGTS objects, and thus we exclude objects with missing photometry. This reduces our sample of 8640 objects by 393 to 8247.

### 2.3 Known classified variables in Orion

By identifying known classified variables within our survey, we can form a training set, which can then be used to classify new variables. Previous Orion surveys have identified three common classes of low and high amplitude variables: dippers, EBs, and spotty periodic stars. We cross-matched the classified variables to the NGTS sources using a positional crossmatch with radius 2 arcseconds, using AS-TROPY (Astropy Collaboration et al. 2013, Astropy Collaboration et al. 2018), finding:

- a total of 17 dipper candidates: 6 from Morales-Calderón et al. (2011), and 11 AA Tau analogs from Rice et al. (2015), which we considered to be dippers,

- 174 periodic candidates from Rice et al. (2015),

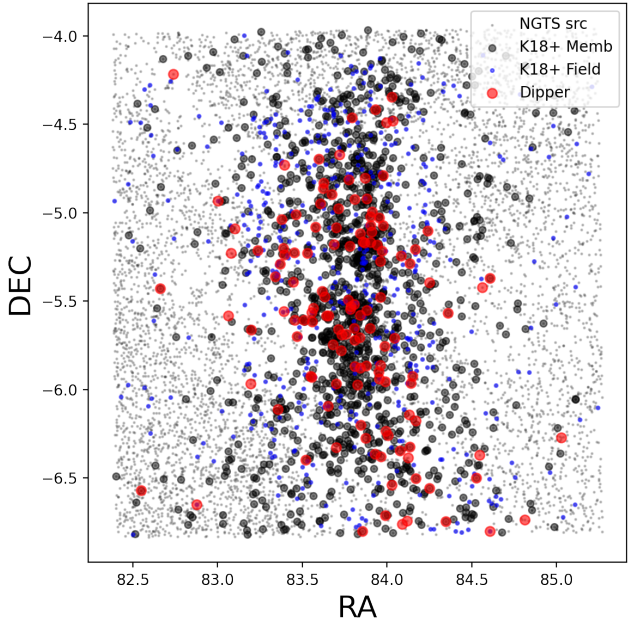

**Figure 1.** The spatial distribution of all NGTS sources in the ONC survey field (shown as the smallest black symbols). Objects in the K18 sample as confirmed or uncertain members are shown with the larger black circles. Objects identified by K18 as field stars are marked with blue symbols, and the final sample of NGTS Orion dipper stars are marked in red.

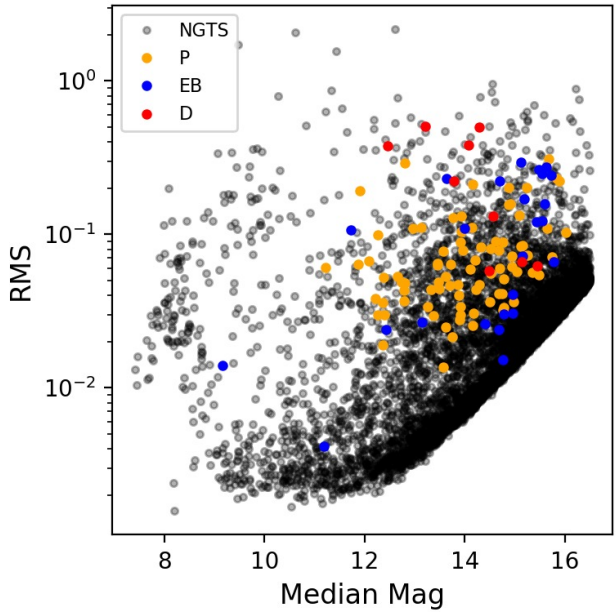

**Figure 2.** Plot of robust RMS versus median magnitude for the final set of 8247 sources in NGTS Orion survey (using binned lightcurves). We overlay the classified objects with coloured symbols (dippers: red, EBs: blue, periodic variables: orange). The RMS is calculated as $1.4826 \times$ MAD (Median Absolute Deviation) of the lightcurve magnitudes, and is less sensitive to outliers than the standard deviation (see e.g. Rousseeuw & Croux 1993).

• a total of 44 EB candidates: 3 from Rice et al. (2015), 28 from Simbad (Wenger et al. 2000), and a further 13 flagged by the NGTS planet search (Wheatley et al. 2018).

Our decision to consider AA Tau analogs as dippers was based on the dipper-like appearance of several of these objects in the NGTS data and ambiguities that are present in current definitions of AA Tau analogs vs. dippers (Cody et al. 2014; Rice et al. 2015; Roggero et al. 2021).

We vetted the binned NGTS lightcurves of each object in the training set to ensure that each source conformed to our expectations for the respective class. The final cross-matched data set consisted of 9 dippers, as well as 25 EBs and 85 periodic variables (see Table 1). NGTS lightcurves for the 9 dippers are presented in Fig. **??**. All the known classified variables are highlighted in Fig. 3.

## 3 IDENTIFICATION OF NEW VARIABLE STARS

We used the NGTS lightcurves to split the sample into variable and non-variable sources. For quiet stars, we see that the magnitude RMS (root-mean-square deviation about the median) of the lightcurves shows a clear sequence of quiet stars (see Fig. 2), and a distinct population spread above this sequence (supported by the location of the classified variables). We found that the RMS for quiet stars scales linearly with the average of the errors in the re-binned NGTS lightcurves, i.e. $\mathrm{RMS_{pred}} = \mathrm{err_{rebin}} \times 1.4 + 0.002$ magnitudes, which means that we can infer what the *RMS* should be for a quiet star based on the average of the photometric errors in its lightcurve. Thus a metric can be employed which estimates the difference between the measured photometric noise of a source, and our expectation for that source if it were not variable. In Figure 3 we plot this metric ($RMS - RMS_{\mathrm{pred}}$) as a function of magnitude, and again overlay the previously classified variables.

Dippers are high amplitude variables, and so we applied a threshold of 0.01 for the distinction between a variable and a non-variable NGTS source (see the bottom panel of Fig. 3). In Fig. 3 nearly all the classified variables sit above our 1 per cent threshold. We have excluded three previously classified variables: one periodic star, and two EBs. Inspection of the NGTS lightcurves for these three objects shows that both EBs are obviously variable (albeit with a low duty cycle, in that the eclipse happens over a small fraction of the orbit), while the periodic variable has a very low amplitude. Thus we caution that our $RMS - RMS_{\mathrm{pred}}$ metric will likely have excluded some low-amplitude dippers, periodics and EBs, and at the faint end will include some stars which are not intrinsically variable.

We also note the fraction of stars that are variable as a function of magnitude via the histogram in the top panel of Fig. 3, which is around 25 per cent for stars fainter than mag=9. We also see a significant increase in the fraction of variable stars brighter than mag=9, which is due to saturation of the CCD combined with variable observing conditions (seeing and sky transparency). Following the application $RMS - RMS_{\mathrm{pred}} >= 0.01$, we reduce our sample of 8247 objects in Orion with clean data and complete photometry, to 2102 stellar variables, of which K18 previously categorized 943 as Orion members and 105 as field stars. In this variable set, 119 were previously known and classified as either periodic rotators, eclipsing binaries, or dippers. Finally, we also flag that a negligible proportion of our objects with magnitude brighter than 10 exhibit saturation features. We vetted the entirety of the previously classified objects to ensure the training set's purity.

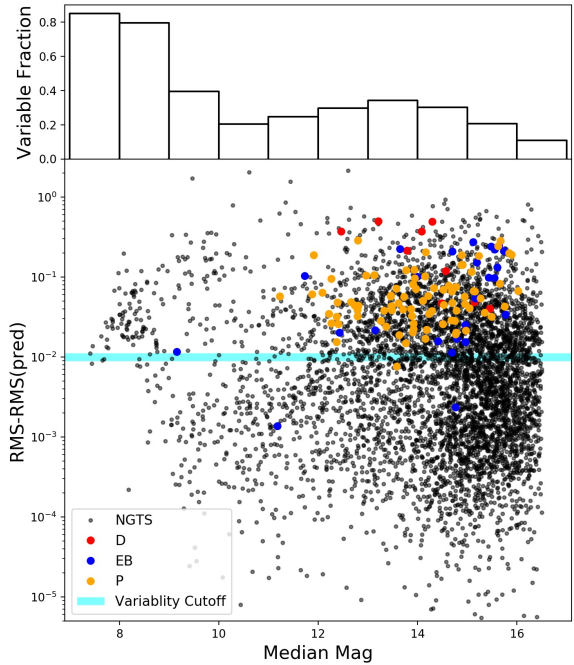

**Figure 3.** Plot of the difference between the observed RMS and the predicted RMS for a quiet star as a function of NGTS median magnitude (bottom) as well as the fraction of objects in each magnitude bin which pass the variability criterion (top). All 8247 NGTS sources which survive quality control are shown in the figure (as black dots). Also shown are the NGTS parameters for previously classified variables (red: dippers, blue: EBs, orange: periodics). The defined cut off for variability at 1 per cent is shown as a horizontal blue line.

## 4 CLASSIFICATION OF STELLAR VARIABLES: RANDOM FOREST

Our primary goal was to identify and validate the dipper population in Orion, which necessitated the differentiation of dippers from the most common classes of variable stars seen in the cluster, such as EB systems or spotty stars. Machine learning classification techniques have been shown effective in classifying stellar variables (e.g. [Richards et al. 2011](#), [Bloom et al. 2012](#), [Mackenzie et al. 2016](#)). In particular, the ensemble machine learning classification algorithm Random Forest (RF) has shown to be robust in distinguishing dippers from other stellar variables ([Hedges et al. 2018](#)). The RF classification algorithm relies on a series of individual decision trees which take as input quantitative features of a labelled population in order to characterize the population characteristics of each label or class. Individual decision trees then consider the features of new, unclassified objects in randomized order, and assign a probability that the new object belongs to any individual class based on the values of the features that the tree will essentially branch through. RF classifiers then average the class probabilities from each tree in order to make a final prediction on the probability an object belongs to each class - the highest valued probability is therefore the classification assigned by RF. RF has extensive astrophysical applications beyond stellar classification, including supernova classification (e.g. [Bailey et al.](#)

[2007](#)) and planet identification (e.g. [Henghes et al. 2021](#)). For more information on the technical details of RF, see [Breiman](#) ([2001](#)).

Like all machine learning classification algorithms, our RF classifier required a labelled training set containing features (i.e. quantitative properties of the NGTS stars, mostly derived from their lightcurves). Supervised machine learned algorithms like RF can only classify objects according to the known classes represented in its training data. The training set is composed of those 119 sources previously flagged and now vetted as confirmed periodics, dippers, and EBs in Orion. In section 4.1, we detail the features used in our classifier for stellar characterization. In section 4.2, we describe the implementation and validation of our machine learned classifier. In section 4.3, we outline our method for identifying which of the dippers show periodic behaviour, and in section 4.4, we introduce a method for the identification of disk-bearing objects.

### 4.1 Features for Variable Classification

The features implemented in our machine learning classification scheme were informed by previous approaches for stellar lightcurve classification (e.g. [Richards et al. 2011](#) and [Hedges et al. 2018](#)) and previous studies of dipper stars (e.g. [Cody et al. 2014](#)). We categorized our features according to 3 main areas: statistical properties of the lightcurves, periodic analyses (Lomb-Scargle, Box Least Squares), and infrared photometry indicative of the presence of a circumstellar disk. All 14 features used in our classifier are summarized in Table [2](#).

#### 4.1.1 Statistical Lightcurve Features

We used 6 features to capture the statistical properties of our lightcurves (the first 6 in Table [2](#)).

Flux asymmetry $M$, and the aperiodicity $Q$, have been shown to be particularly useful for dipper classification ([Cody et al. 2014](#), [Hedges et al. 2018](#) and [Bredall et al. 2020](#)).

$M$ measures the directional asymmetry of the lightcurve flux:

$$M = \frac{d_{10} - d_{med}}{\sigma} \tag{1}$$

where $d_{10}$ is the mean of all flux values in the top and bottom deciles of the lightcurve, $d_{med}$ is the median magnitude value of the entire lightcurve, and $\sigma$ is the overall rms of the lightcurve.

The $Q$ statistic measures the lightcurve aperiodicity, with a score of 0 describing a periodic variable and a score of 1 describing a completely stochastic variable. First we derived a phase-folded lightcurve (using the Lomb-Scargle period) which was binned in phase space by taking the median magnitude in each of 250 bins. A residual lightcurve is built by subtracting this binned lightcurve from the raw lightcurve at each point in time (equivalently phase). $Q$ is the ratio of the RMS of the residual lightcurve to the original lightcurve, i.e.:

$$Q = \frac{RMS_{residual}}{RMS_{original}} \tag{2}$$

An example of the process for deriving the residual lightcurve is given in Fig. [4](#). We show our training set (periodics, EBs, and dippers) in the $M$-$Q$ parameter space in Fig. [5](#), showing the value of these metrics for classification. There is clear overlap between the EBs and Dippers.

The median absolute deviation ($MAD$), the difference between the median and mean magnitudes ($dtav$), and standard deviation were also found to be useful by [Hedges et al.](#) ([2018](#)).

**Table 1.** List of morphology classes fed into the initial RF classifier. The number of sources per morphology class and the references from which the sources were drawn is included.

| Morphology Class | Number of Objects | Sources |
|---|---|---|
| Dippers | 9 | Morales-Calderón et al. 2011, Rice et al. (2015) |
| Eclipsing Binaries | 25 | Morales-Calderón et al. 2011, Rice et al. (2015), NGTS (Wheatley et al. 2018), Simbad (Wenger et al. 2000) |
| Periodics | 85 | Rice et al. (2015) |

**Table 2.** List of lightcurve features used in the RF algorithm to distinguish dipper stars from periodics and EBs.

| Feature | Description |
|---|---|
| $M$ | The flux asymmetry or $M$-statistic defined in Cody et al. (2014). |
| $Q$ | The aperiodicity or $Q$-statistic defined in Cody et al. (2014). |
| $MAD$ | The median absolute deviation of the lightcurve (magnitudes). |
| $dtav$ | The difference between the mean and the median of the lightcurve (magnitudes). |
| $stddev$ | The standard deviation of the lightcurve (magnitudes). |
| $amp9010$ | The difference between the 90th and 10th percentile of the lightcurve (magnitudes). |
| $LSperiod$ | The rotation period found for the lightcurve using the Lomb-Scargle algorithm. |
| $LSpdgrmstddev$ | The standard deviation of the source's Lomb-Scargle periodogram. |
| $LSmaxpow$ | The maximum power value of the Lomb-Scargle periodogram. |
| $BLSperiod$ | The period found for the lightcurve using the Box-Least Squares algorithm. |
| $BLSduration$ | The duration of the lightcurve's transit found for the best-fit transit model determined by BLS. |
| $BLSdepth$ | The depth of the lightcurve transit found for the best-fit transit model determined by BLS. |
| $BLSmaxpow$ | The maximum power value associated with the transit model determined by BLS to be the best fit for the lightcurve. |
| K–W2 | The excess color value taken from subtracting WISE W2 from 2MASS K. |

The amp9010 metric, was inspired by a combination of the "diff" metric used in Hedges et al. (2018) and the $\delta_{90}$ metric used in Rodriguez et al. (2017). amp9010 is less sensitive to systematic outliers than these other metrics, but encodes similar ideas.

#### 4.1.2 Periodic Features Derived from Lomb-Scargle

Lomb-Scargle periodograms have been previously used to identify rotation periods in stellar clusters (e.g. Rebull et al. 2016, Gillen et al. 2020a). The Lomb-Scargle algorithm for detecting and characterizing periodic signals in unevenly sampled data is the optimal statistic for fitting a sinusoid to data in the presence of Gaussian noise – under these assumptions, stellar rotation periods can be estimated (for more information on Lomb-Scargle, see VanderPlas 2018). As Gillen et al. (2020a) note and validate via their work in Blanco 1, the assumption of a constant period and non-modulating lightcurve is reasonable for the low-mass stars which are the sample of this study.

We used the ASTROPY implementation of Lomb-Scargle, applying a logarhithmic grid of 4000 frequencies between $1/30 \ day^{-1}$ and $12 \ day^{-1}$, and extracted the following features: the maximum power of the periodogram, the period at this peak, and the standard deviation of the periodogram. This power, which is between 0 and 1, can be viewed as the algorithm's confidence that the selected period is correct.

We note that two non-classified variable objects failed to produce finite values when run through the LS pipeline. We chose to omit these variables from further analysis, reducing the number of variables to 2100.

#### 4.1.3 Periodic Features Derived From Box-Least Squares

While Lomb-Scargle is appropriate for identifying sinusoidal signals in periodic variables, it does not perform so well for non-sinusoidal signals - like those of EBs and planetary transits. Algorithms like Box-Least Squares, developed by Kovács et al. (2002), are better suited for capturing the box-like nature of these transiting events. This method has been used previously to distinguish EB signals from typical periodic signals (e.g. Dékány & Kovács (2009)).

The benefit of including BLS metrics in our classifier stems from its advantage over Lomb-Scargle in classifying EB periods. BLS provides a better model for the lightcurve of an EB than does a sinusoid, as demonstrated in Fig. 6.

We used the ASTROPY implementation of BLS, applying a logarthimic grid of 20000 periods between 0.1 day and 30 days as well as a logarithmic grid of 50 transit durations between .01 days and .099 days. We included four BLS metrics in our classifier: period, maximum power, transit depth, and transit duration.

#### 4.1.4 A feature based on the presence of a disk

Because the association between dippers and circumstellar disks has been strongly established, we added K-W2, a strong disk indicator (e.g Yao et al. 2018 hereafter Yao18, Luhman & Mamajek 2012), as an additional feature. As an alternative approach, Hedges et al. (2018) used WISE colors (associated with the presence of a circumstellar disk) to validate their machine. We found (outlined in Appendix A) that the K-W2 metric was the most valuable feature for the disk identification out of 10 magnitudes and colors incorporating NGTS, Gaia, 2MASS, and WISE photometry. The K and W2 bands have effective wavelengths of ~2.16 and ~4.62 microns respectively.

### 4.2 Implementing and Validating an Iterative Random Forest

The variables in the ONC appear to be dominated by three main classes (at least in the magnitude-, amplitude- and time-range we have explored with NGTS): periodics, EBs and dippers. There are of course other broad and overlapping classifications (see Herbst 2012 for an overview). But for this exercise, and to meet our primary goal, we focused on the identification of these first three classes.

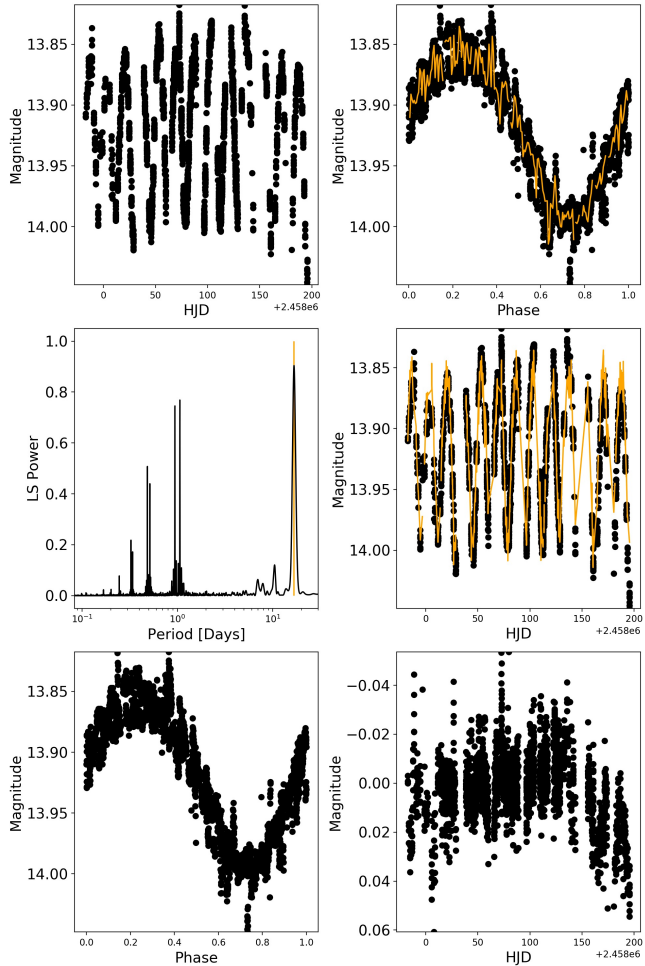

**Figure 4.** Visualization of the steps taken to derive a stellar period and aperiodicity value, which are outlined in the text, taken for example object NGTS 16798, a periodic YSO used in our training set. Starting from top to bottom on the left and then moving from top to bottom on the right: (1) Original NGTS lightcurve, (2) Lomb-Scargle periodogram (note the peak period identified with the orange line), (3) Phase folded lightcurve, (4) phase folded lightcurve (black) with smoothed phase approximation (orange), (5) original lightcurve (black) with full smoothed lightcurve (orange), (6) the result of subtracting the smoothed lightcurve from the original one.

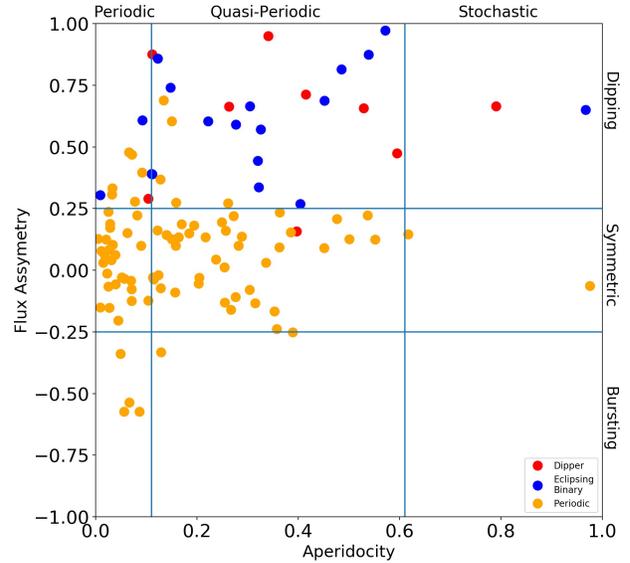

**Figure 5.** Aperiodicity vs Flux Asymmetry for the Training Set Objects. The demarcations made along each axis are taken from Cody et al. (2014), who used them to differentiate stars by morphology class. Note the clustering of periodics towards $M = 0$ (but with some spread in $Q$), and the clustering of dippers towards high $M$ and a uniform spread in $Q$. The overlap of the dipper population with the EB population suggests that the Lomb-Scargle period may not be robust for EBs. We used the Lomb-Scargle period to derive $Q$ in order to normalize it across classes.

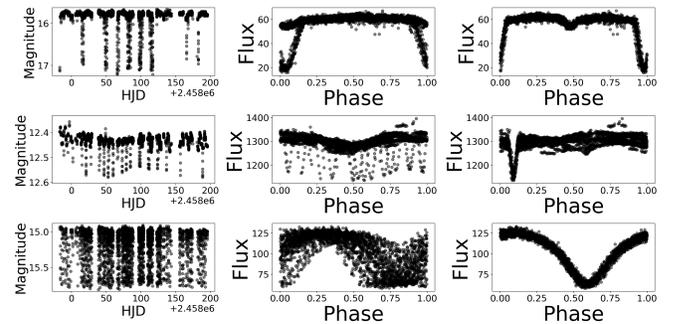

**Figure 6.** NGTS lightcurves for three training set EBs (object IDs from top to bottom: 5916, 21306, 716) all identified in previous surveys. The first panel is the entire lightcurve plotted over the observing period, the second is phase folded to the Lomb-Scargle period, the third is phase-folded to the BLS period. Note the improvement in all cases.

With the prior construction and vetting of our labelled training set, and the development of quantitative features able to describe our labelled populations and our unclassified stars, we proceeded with the implementation of a RF machine learning algorithm to classify the bulk population of NGTS variable stars. Our sample was composed of the 1986 objects identified as new ONC variables (Section 3), together with a training set of 119 classified variables (9 Dippers, 25 EBs, 85 periodics: see Section 2.3).

We implemented RF in the SKLEARN environment (Pedregosa et al. 2011), with a train-test-split of 80-20. That is, 20 per cent of the stars in our training set were removed in each run of RF so that the classification mechanism could be evaluated by identifying whether or not training set stars were correctly classified by the machine (see Figure 7 for an example).

We used RF in a two-step, iterative process. This decision was made for two reasons: firstly, due to the relatively small dipper pop-ulation in our training set, the initial classifier was likely limited in its ability to robustly classify new dippers. So, we implemented a first step of RF classification for which we visually confirmed the machine results for those dippers with probabilities greater than 0.8 (via inspection of lightcurves, and postage stamp NGTS images of the source). Out of 40 candidate dippers with P>0.8, 3 were rejected as being artefacts. Thus for the second round the training set for Dippers was increased to 46 objects.

We also added 4 newly discovered, visually vetted EBs to our training set following visual inspection of the all P>0.5 EB candidates (phase folded to their box-least squares periods). We did not add in the newly discovered periodic candidates to the training set, given that this population was already strongly represented.

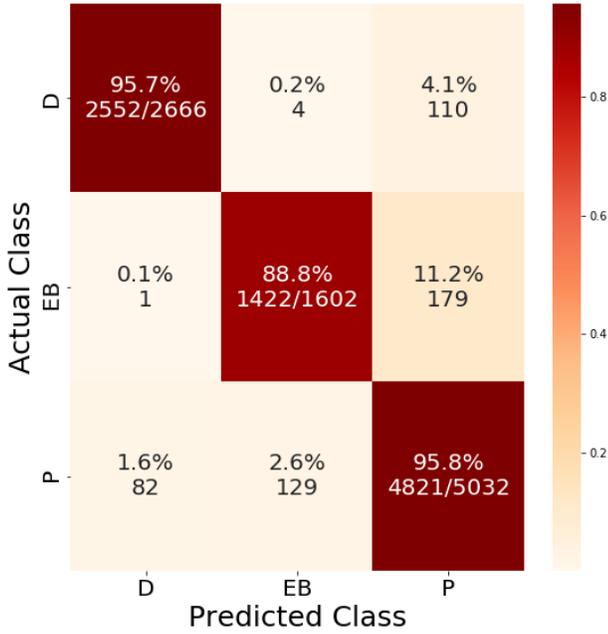

**Figure 7.** Confusion matrix for the second round classifier after 300 iterations using a 80-20 train-test-split.

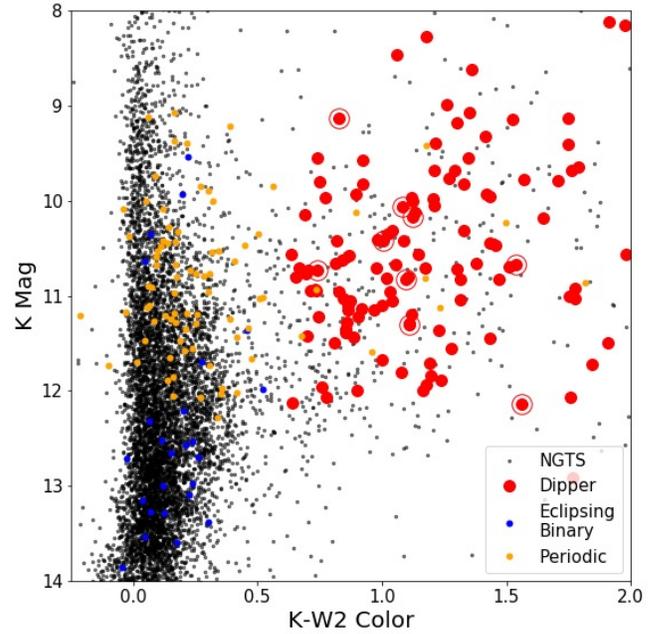

**Figure 8.** 2MASS-WISE K-KW2 CMD for the final set of classified sources following two-step RF and visual vetting, indicated by class. The 9 Dippers enclosed by a red circle are from the original training set and all 8247 NGTS sources which survive quality control are shown in the figure (as black dots)

We then ran our second round of RF to improve our classifications, particularly for the dippers. The accuracy of our two-step classification mechanism can be seen in the second round confusion matrix, shown in Fig. 7. The matrix is strongly diagonalized, because of the machine's strong performance in classifying known members of each class when they are separated via train-test-split. It is especially relevant to note the improvement of dipper recovery from rounds 1 to 2; in round 1 dipper recovery is 71.7 percent, and in round 2 it improves to 95.7 per cent. This speaks to the reliance of the machine on larger training set populations to properly classify new sources. Still, the performance in both rounds is strong, and the second round performance especially validates the machine's ability to correctly identify dippers. As well as increasing the dipper sample, this iteration improved the machine's EB recovery from 74.1 per cent to 88.8 per cent in the second iteration.

We vetted (eyeballing of lightcurves and images) all dippers and EBs with a probability greater than 0.5 discovered in the second round. This added 83 out of a total of 151 candidate dippers. The periodics were not vetted in the same way, but the probabilities in the data table can be used to refine the sample.

In both rounds of RF, we set the number of trees to 50, and the number of features used by the classifier was set to be equal to the total number of features described in Table 2 (14). We ran the RF algorithm 300 times on the sample in both classification rounds. This helped to compensate for the way a train-test-split would strongly impact the population characteristics for different classes, especially dippers in the first round of the machine. At this stage, there would have been an expected 2–3 dippers used to validate the performance of the machine and only 6–7 used for classification. By running RF 300 times, we ensured as many potential combinations of training sets were included in the machine, such that our machine was less biased with respect to the random outcome of the train-test-split.

RF also measures the relative importance of each feature in making its classification, where the relative importance scores of all features sum to one. We show our importance scores for the fourteen features (derived from the second round of the classifier) in Fig. 9. These scores shed light on the metrics which were most useful in classifying these objects. The flux asymmetry and K-W2 metrics can be easily linked to dipper classification - this can be seen in the 2MASS-WISE K vs K-W2 CMD when compared to the final sample (Fig. 8). The importance of the flux asymmetry supports the use of the feature by Cody et al. (2014), Bredall et al. (2020) and others in relying on this feature in dipper identification. The importance of the IR excess metric, as is seen in both its importance score as well as in how the dippers cluster at larger IR excess in Fig. 10 shows the important connection between disks and the dipping phenomenon.

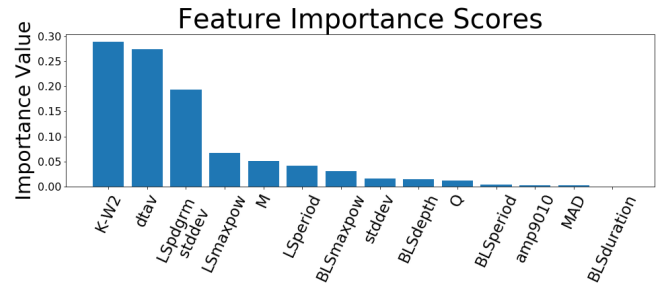

**Figure 9.** Relative importance assigned to each feature in the second round RF classifier.

The high importance of certain Lomb-Scargle and BLS metrics points to their effectiveness in classifying periodics and EBs, respectively. The parameter space occupied by the 7 most important features (as determined by the second round classifier) for the final set of candidates is shown in Fig. 10. The *dtav* metric is the second most important feature, and behaves similarly to *M*, but appears to have power when *M* 'saturates' (turns over). We see a strong cor-relation between the $LSpdrgmstddev$ and $LSmaxpower$ metric, particularly with high values for the periodic objects. Otherwise, we do not observe strong correlations between our important features across each class, important for ensuring that no groups of variables induce a strong bias in the machine's selection.

### 4.3 Classified variables

We computed the results of our classification algorithm for 2100 variable stars (full table available online: column headings in Table 3). A total of 942 out of 2100 variables were classified by the machine (and vetted in the case of EBs and dippers). The remaining 1160 never achieved a probability>0.5 for a single class, or failed vetting for the dipper and EB classes. We show the classified and unclassified variables in a series of colour-magnitude diagrams in Fig. 11, compared to confirmed and possible members of the ONC (K18), as well as the backdrop of all NGTS ONC targets. The periodics and dippers adhere very closely to the cluster sequence, as we would expect. The unclassified variables show a weaker correlation, however there are still large numbers of unclassified variables which appear to be in the ONC.

Our total dipper population is 129 objects (of which 9 were in the training set). The total number of new EBs was 4, and the final number of new periodic variable candidates for P>0.90 was 421, although there are clear periodic signatures in objects with probabilities as low as P>0.5.

These candidate EBs are NGTS objects 1226, 22852, 33482, and 1519, and all lie within the periphery of the NGTS field and away from the centre of the ONC. Their Gaia eDR3 parallaxes and proper motions indicate that they are unlikely to be members of Orion, but rather more distant objects. Furthermore, their lightcurves do not exhibit obvious characteristics of young stars. Nevertheless, we attempted to jointly model the NGTS lightcurve and system SED (constructed from available broadband photometry) for NGTS 1226 (Gaia eDR3 source_id: 3215666723464141568) and NGTS 22852 (Gaia eDR3 source_id: 3017105471927497728), but large uncertainties on the parallaxes meant that we could not constrain the distances or radii in either case. Additionally, NGTS 1226 has a large re-normalised unit weight error (RUWE = 4.54), suggestive of perturbed astrometry and a possible tertiary companion (Stassun & Torres 2021). NGTS 33482 (Gaia eDR3 source_id: 3016929756224963072), a likely contact binary, and EB 1519 (Gaia eDR3 source_id: 3209887239375934848), the most tenuous identification, are left for future study.

### 4.4 Determination of Periodic Variability in Dippers

Space-based studies of dippers have shown the importance of periodicity in the lightcurves for a significant fraction of dippers. Roggero et al. (2021) used K2 observations in Taurus to identify 34 dippers for which they split their lightcurves evenly between aperiodic and quasi-periodic. Ansdell et al. (2016) also found a 50:50 split in the sample of 10 dippers discovered in Upper Scorpius with K2. Hedges et al. (2018), Ansdell et al. (2018) and Roggero et al. (2021) all used (quasi-)periodic dippers to attempt to gain insights on the location of the material causing the dips. In this paper we have attempted to search for periodicity in our sample of Orion dippers, although more challenging with ground-based data.

We re-ran the generalised Lomb-Scargle periodogram for all dipper stars in our sample, but applied additional methods to obtain more robust periods (see e.g. VanderPlas 2018). To begin with we used the same frequency limits as in Section 4.1.2, but increased the resolution of the grid by a factor of ten. Next, we used this frequency grid to compute the window function of our sample following the method outlined in VanderPlas (2018). An example window function is visualized in Fig. 13. We do see strong alias peaks at $P/(1 \pm n * P)$ days where n is an integer, a typical feature of window functions in ground-based surveys, illustrating that these periods that should not be trusted.

For 13 of our dippers, the Lomb-Scargle periodograms showed an anomalously high baseline. We found that the lightcurves of all 13 of these objects contained large outlier datapoints, i.e. unrealistically bright and/or dim single measurements. We implemented a simple 3-sigma data cleaning step to remove these outliers from all lightcurves, and recalculated the periodograms.

We considered the three highest peaks in each periodogram which met two criteria: (1) the selected peak must correspond to a period separated by at least 0.1 day from other selected periods, and (2) the selected peak must correspond to a period not within 0.1/n days of a 1/n day alias period. Previous studies have applied an FAP threshold of 1-5 per cent for validation of stellar period peaks (eg Cargile et al. 2014, Reinhold et al. 2017), the periodogram peak powers for our objects derived with the Baluev method - which provides an upper bound of the FAP relative to other methods and therefore provides a conservative estimate of the FAP VanderPlas 2018 - are consistently 10s to 100s order of magnitudes smaller than recommended literature thresholds. Thus we opted to omit this step from our analysis, but note that these small FAP values would have been accepted in previous studies.

Following the computation of the three highest distinct, non-aliased periodogram peaks, we visually examined each object's lightcurve, periodogram, and the three phase-folded lightcurves corresponding to the object's three peak periods corresponding to the highest periodogram peaks.

We then assigned to each object a final periodicity status which crudely resembles the classification scheme outlined in Newton et al. (2016), where each dipper lightcurve the status of "secure period" (SP), "tentative period" (TP), or "no period" (NP). In Fig. 14, we show the Lomb-Scargle peak powers versus aperiodicity $Q$ for all dippers, where the $Q$ has been recomputed at the revised periods. Although there is a large spread in $LSmaxpow$ at any given $Q$, the SP objects appear correlated, and the NP objects sit systematically below. Therefore our final sample of rotators fit one of two categories: all SP objects regardless of Lomb-Scargle power, and all TP objects which satisfy the criteria $LSmaxpow > 0.2$ (the SP/TP nomenclature is retained however for further analysis considerations).

### 4.5 Disk properties of dippers

As described in Appendix A we classified the disk-bearing nature of our entire survey, using a machine learning approach, and dominated by the K–W2 infrared excess ((2MASS, WISE). Out of 129 dippers, 128 were found to bear disks. It is worth noting that all of our training set dippers bear disks, and furthermore K-W2 is the single most important feature for both the identification of a dipper, and the presence of a disk. Thus we have introduced a bias here, specifically against diskless stars which are also dippers. Indeed it seems more likely that the lone apparently diskless dipper in our discovered sample, is incorrectly classified as not having a disk.

Firstly, while the machine's strength at disk-bearing classification is evident from Fig. A2, the machine is still 9.4 per cent inaccurate in identifying disk-bearing objects. Additionally, it may be that disk classification at even longer wavelengths may work better in this instance (see e.g. Luhman & Mamajek 2012. Also the star is a con-firmed ONC member (K18), although this in itself is not completely dependable, given that of the 128 confirmed disk-bearing dippers, 10 are classified by K18 as field stars and 17 are classified with unknown membership.

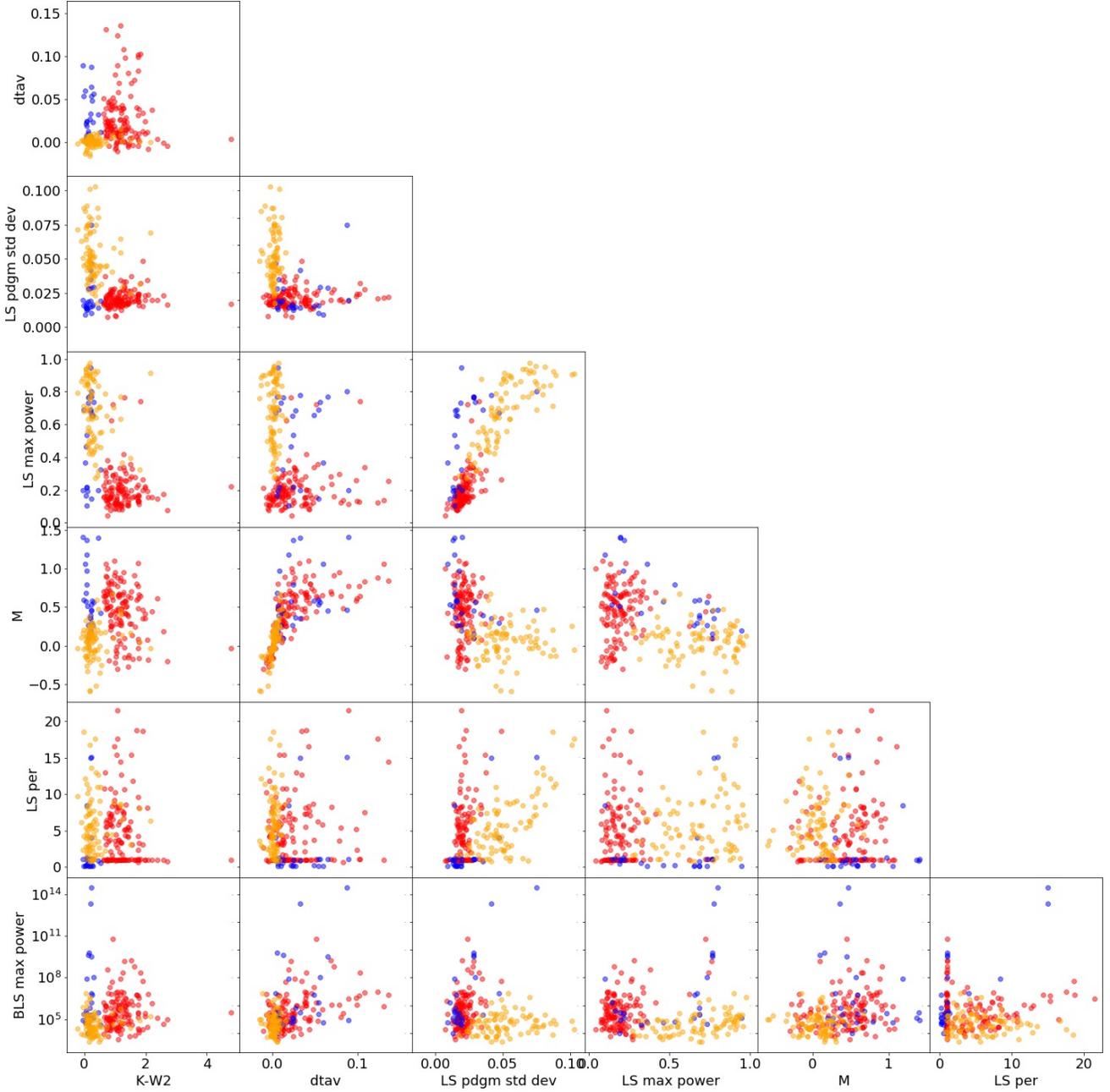

**Figure 10.** Corner plot showing key features for training set dippers (red), EBs (blue), and periodics (orange). We also include newly discovered EBs and dippers in the figure (but do not include newly classified periodics).

In the end we chose to treat the entire sample of 129 dippers as disk-bearing ONC members. We also note here that three of our dippers were independently discovered in TESS (Capistrant et al., in prep), supporting our identification. These objects are NGTS 11441 (TIC 11284400), NGTS 32106 (TIC 332969781), and NGTS 24392 (TIC 276664304).

## 5 NEW ORION DIPPERS

### 5.1 Spectral types of dippers

Dippers have been primarily identified as K and M type stars (e.g. Cody et al. 2014, Hedges et al. 2018). This differentiates them from larger-mass YSOs with similar lightcurve morphology such as UX Orionis (as discussed in Bredall et al. 2020). For our sample we have temperature data on 1372 stars from K18: 336 of which are disk-bearing objects and a further 76 are dippers. We map these stellar temperatures onto spectral type following the quantitative relationship constructed for YSOs by Pecaut & Mamajek (2013). A histogram displaying the absolute and relative frequencies of all stars, disk-bearing objects, and dippers by spectral type is presented in Fig. 15.

**Table 3.** Column headings for the table of data for all NGTS variables. The table is available in its entirety in machine-readable form.

| Number | Column | Contents |
|---|---|---|
| 1 | Index | Running number (0 indexed) |
| 2 | NGTS ID | NGTS object identification (cycle 1807 pipeline run) |
| 3 | RA | NGTS Right ascension (J2000) |
| 4 | DEC | NGTS Declination (J2000) |
| 5 | NGTS MEAN MAG | NGTS Mean magnitude |
| 6 | NGTS MAG RMS | NGTS lightcurve RMS (computed as 1.48x(Median Absolute Deviation) |
| 7 | Flux asymmetry | (or M-statistic), see Section 4.1.1 |
| 8 | Lomb-Scargle period | in days |
| 9 | LS pdgm std dev | Standard deviation of the Lomb-Scargle periodogram |
| 10 | LS MAX Power | Maximum power of the Lomb-Scargle periodogram |
| 11 | Aperiodicity Q | Aperiodicity of the lightcurve (Q-statistic), see Section 4.1.1 |
| 12 | amp9010 | Difference between the 90th and 10th percentile of the lightcurve (magnitudes) |
| 13 | std dev | Standard deviation of the lightcurve (magnitudes) |
| 14 | MAD | Median absolute deviation of the lightcurve (magnitudes) |
| 15 | dtav | Difference between the mean and the median of the lightcurve (magnitudes) |
| 16 | BLS Period | Box least squares period |
| 17 | BLS Max power | Maximum power of the BLS periodogram |
| 18 | BLS duration | Duration of the transit (days) |
| 19 | BLS depth | Depth of the transit (magnitudes) |
| 20 | K-W2 | K(2MASS)-W2(WISE) (magnitudes) |
| 21 | Gaia G | Gaia G from eDR3 (magnitudes) |
| 22 | Gaia BP | Gaia BP from eDR3 (magnitudes) |
| 23 | Gaia RP | Gaia RP from eDR3 (magnitudes) |
| 24 | 2MASS J | 2MASS J-band (magnitudes) |
| 25 | 2MASS H | 2MASS H-band (magnitudes) |
| 26 | 2MASS K | 2MASS Ks-band (magnitudes) |
| 27 | WISE W1 | WISE W1-band (magnitudes) |
| 28 | WISE W2 | WISE W2-band (magnitudes) |
| 29 | WISE W3 | WISE W3-band (magnitudes) |
| 30 | WISE W4 | WISE W4-band (magnitudes) |
| 31 | P Probability | Periodic probability |
| 32 | D Probability | Dipper probability |
| 33 | EB Probability | Eclipsing binary probability |
| 34 | Class | Assigned class |
| 35 | Original Classification Source | Source of training set classification |

We see that roughly 65 percent of the 1372 member stars are K and M type. Disk-bearing objects and dipper stars favor later K and M spectral types more strongly than the overall population. It is notable that 2 objects classified as dippers are found to be earlier G to F type stars. Inspection of these objects shows that they exhibit dips which are longer in duration and occur more infrequently. This is suggestive that they may instead be a related class of Type III Variable such as a UX Orionis type star.

## 5.2 Disk Evolution of Dippers

Previous works to characterize dippers have noted that these stars bear disks at the early stage of disk evolution. Hedges et al. (2018) found this to be true for Upper Sco and Rho Oph, and Bredall et al. (2020) confirmed the same in Lupus. Bredall et al. (2020) likewise identified this finding for dippers previously found in Upper Sco and Taurus by Cody & Hillenbrand (2018) and Rodriguez et al. (2017), respectively. We seek to discover whether the Orion Dipper population conforms to these previous findings.

Luhman & Mamajek (2012) developed a series of metrics incorporating 2MASS and WISE photometry to identify disk evolutionary states for stars in Upper Sco. Beginning with the earliest disk state and moving forward in evolutionary time, stars can bear: full disk, transitional disk, evolved disk, debris disk, or no disk. Earlier stage disks are associated with larger colour excesses in the infrared, i.e. larger values of 2MASS K − WISE W3 and 2MASS K − WISE W4. Their methods align with previous findings that IR excesses are strong indicators of optically thick dust in the presence of a disk (e.g. Lada et al. (2006)).

We identify the disk evolutionary state of our stars based on the implementation carried out in Bredall et al. (2020) following the findings of Luhman & Mamajek (2012). Bredall et al. (2020) divide disks by evolutionary state following the division outlined in Figure 2 of Luhman & Mamajek (2012), specifically in the parameter space of K−W3 and K−W4 (both colours are 2MASS−WISE). Thus we recreate Figure 7 from Bredall et al. (2020) which is shown in Fig. 16.

We see immediately that dippers bear the same early-stage disks as observed in previous studies: full, transitional, or evolved. Notably, the majority of dippers bear full disks, with only 6 bearing evolved and 7 transitional disks. Those with evolved disks bear IR color excesses indicating that they are near the beginning of this evolutionary stage.

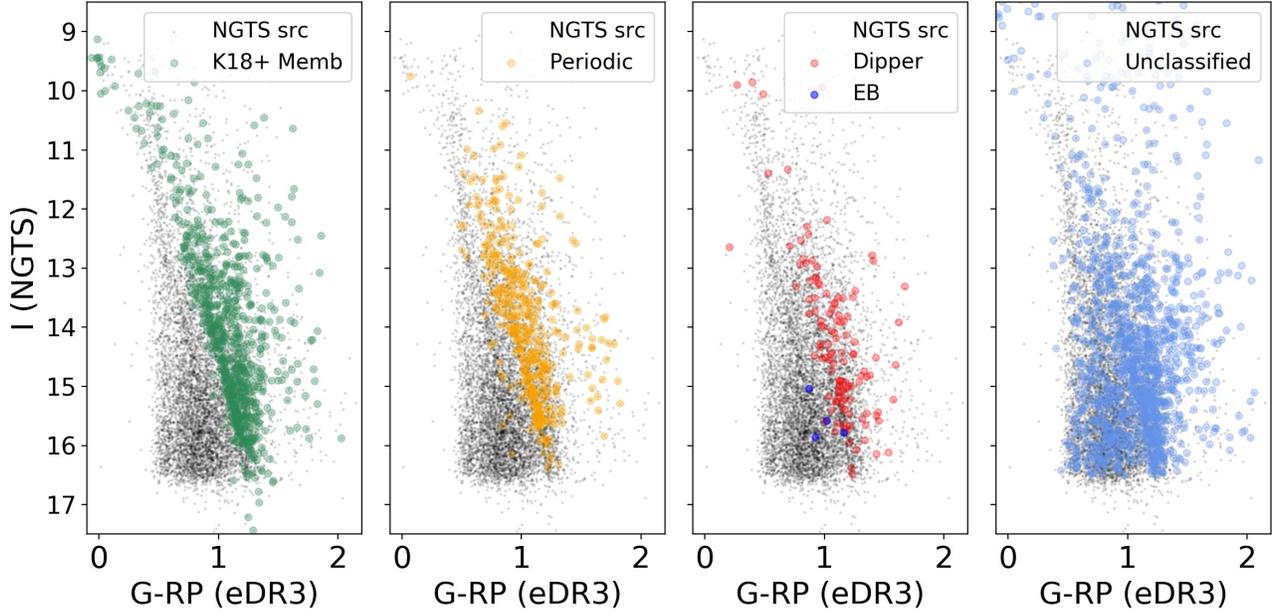

**Figure 11.** G vs G-RP for NGTS sources in the ONC, illustrating differing classes of source. In all 4 panels, black dots denote all the NGTS sources included in the survey with a positional match in Gaia eDR3 (separated by ≤ 2 arcsecond). From left to right, **Panel 1:** green circles are members according to K18, **Panel 2:** Yellow circles are classified as periodic with P>0.9, **Panel 3:** Newly classified dippers (red) and EBs (blue) are overlaid, **Panel 4:** Shows the location of remaining unclassified variable sources with no single class having P>0.5 (light blue).

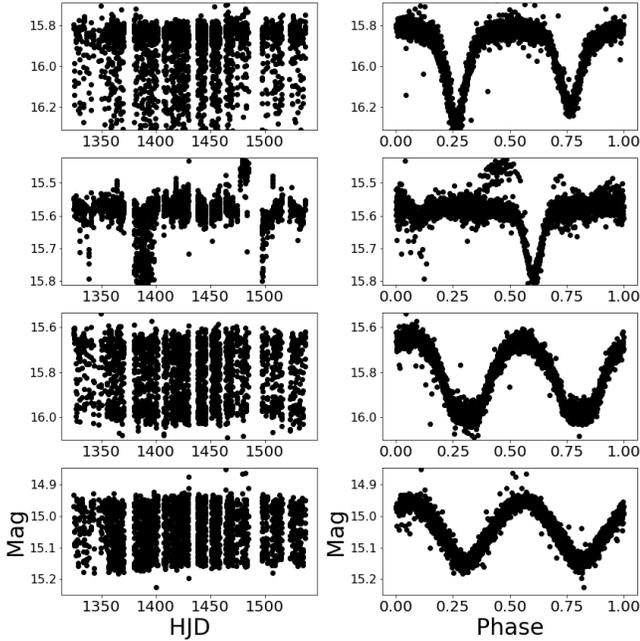

**Figure 12.** NGTS lightcurves for the four EB candidates identified by the stellar morphology RF Classifier with P(EB) > 0.5 which passed visual vetting. From top to bottom, these candidates are NGTS 1226, 22852, 33482, and 1519. lightcurves on the left are plotted in magnitudes and are not phase folded, whereas the curves on the right are plotted in terms of flux and phase folded to the period determined by BLS.

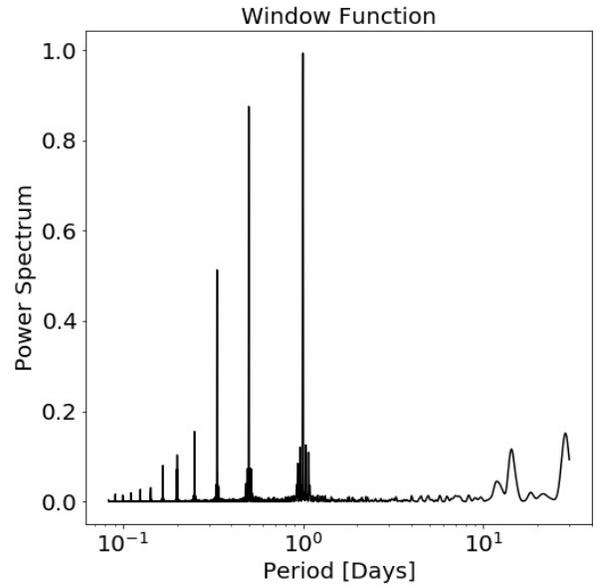

**Figure 13.** The Window Function for the NGTS Orion survey. This window function is computed as an example for dipper object NGTS 12908. We observe alias periods at 1 day, 1/2 days, 1/3 days, 1/4 days.... 1/11 days.

### 5.3 Identifying the Orbital Distance to the Inner Disk Edge

Previous work has established that dippers offer a unique opportunity to study the inner disk region in young low-mass stars (Hedges et al.

2018, Roggero et al. 2021). Here, we used the dipper periods (for both TP and SP objects) to infer the orbital distance between the star and its inner disk edge. By making the assumption that the accretion streams are locked to the inner disk edge (Bodman et al. 2017, Hedges et al. 2018) and that dust in the inner disk orbits the star at a Keplerian velocity, we can assume the orbital period of dust in the inner disk

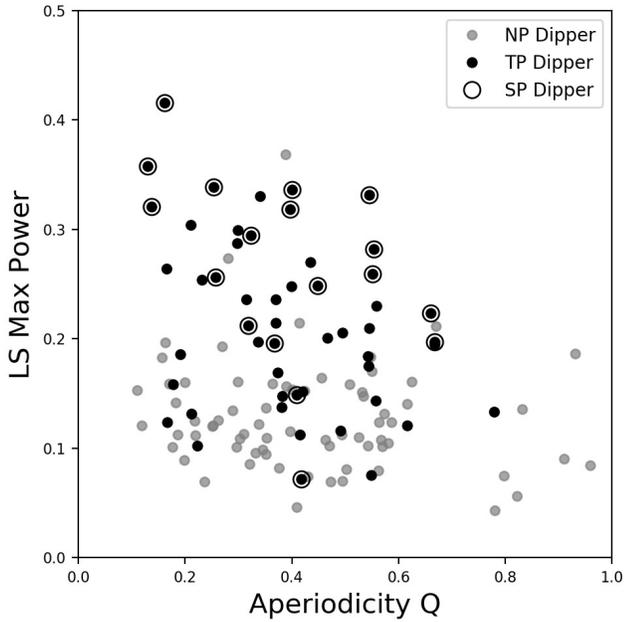

**Figure 14.** Plot of Lomb-Scargle maximum power vs aperiodicity Q for the 129 dippers included in our sample. We show dippers with secure periods (SP), tentative periods (TP), and also those with no reliable measurement of a period (NP).

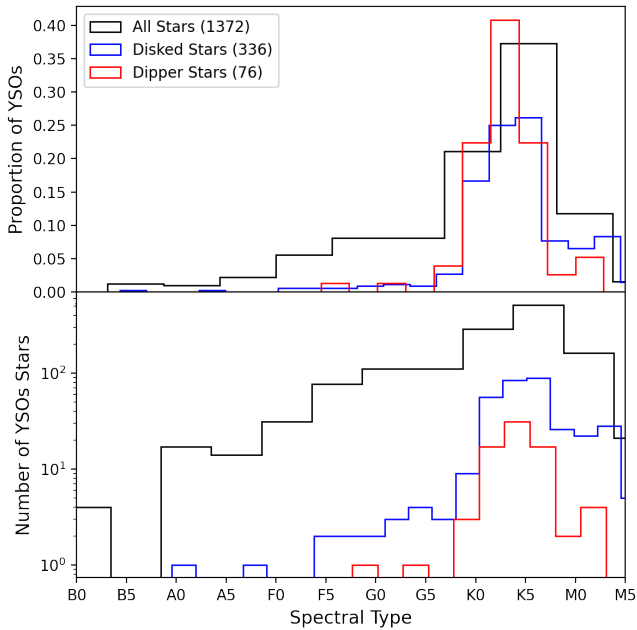

**Figure 15.** Relative, normalized histogram (top) and absolute histogram (bottom) identifying the spectral type spread of all ONC member stars, including disk-bearing members and dipper members. Membership is assigned by K18. Stellar temperature is mapped linearly onto spectral type for young stars following Pecaut & Mamajek (2013).

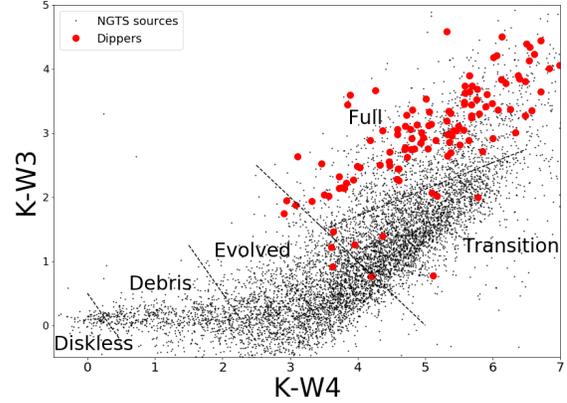

**Figure 16.** 2MASS-WISE color-color diagram. Zones are derived by Bredall et al. (2020) based on the findings of Luhman & Mamajek (2012) regarding circumstellar disk evolution, and are separated by dashed lines. The evolutionary state of the dippers (in red) are consistent with findings on dippers discovered by Ansdell et al. (2016), Rodriguez et al. (2017), Hedges et al. (2018), Cody & Hillenbrand (2018), and Bredall et al. (2020). All 8247 NGTS sources which survive quality control are shown in the figure (as black dots).

edge is given by

$$P = \left(\frac{4\pi^2 r^3}{GM}\right)^{\frac{1}{2}} \tag{3}$$

where $r$ is the orbital distance to inner disk edge, $M$ is the mass of the star, and $G$ is the gravitational constant.

We used spectroscopic $T_{\text{eff}}$ and $\log g$ values for our dippers from Olney et al. (2020) (APOGEE NET) to compare to interpolated stellar models from Baraffe et al. (2015) (hereafter BHAC15). A total of 24 objects have both periods (15 SP, and 9 TP), and spectroscopy. These data, their corresponding errors, and the models are shown in Fig. 17. To derive values and errors for stellar luminosity ($L$) and mass ($M$), we ran Monte-Carlo samples using 1-dimensional gaussians with $\sigma$ equal to the published spectroscopic errors. For each dipper, $L$ and $M$ were assigned to the medians of the resulting distributions, and their errors were set as the 16th and 84th percentiles. The dippers can be seen to cluster between the 0.9 and $\sim 0.2 M_{\odot}$ tracks.

We also attempted to derive effective temperatures and radii for the dippers by modelling spectral energy distributions with both BT-Settl and Pheonix stellar atmosphere models (Allard et al. 2012; Husser et al. 2013), following the method presented in (Gillen et al. 2017, 2020b; Smith et al. 2021). However, we were unable to break the degeneracy between temperature and reddening in Orion. We found that the models typically preferred lower reddening, and hence lower effective temperatures, than estimated from spectroscopic methods. Possible reasons for this include differences between observed SEDs of young stars and existing atmosphere models, or perhaps that the extinction law in Orion does not closely follow the average for the galaxy. While the latter is likely true to some extent, we suspect the former is more more important here.

For the 24 dippers, we show both the period (Fig. 18) and the distance to inner disk edge (Fig. 19) as a function of $L$. To both figures we add in the values computed by Hedges et al. (2018).

Theoretically, the orbital distance to inner disk edge $r$ also equates to the sublimation radius, within which dust will sublimate due to heat from the star. For a given sublimation temperature $T_{sub}$, stellar luminosity L and assuming black body behavior, Hedges et al. (2018)

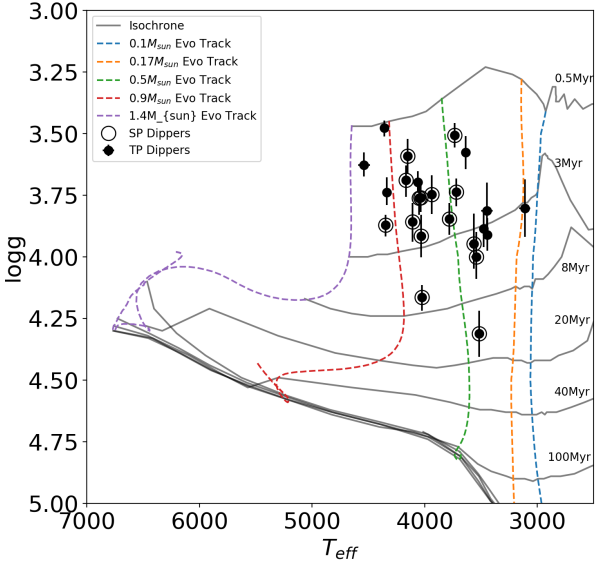

**Figure 17.** $\log g$ vs $T_{\text{eff}}$ for the Orion dippers with tentative and secure periods. Isochrones from Baraffe et al. (2015) are overlaid. We interpolate our dippers across these isochrones to obtain estimates for the masses, luminosities, radii. We make no assumption on the age of each dipper in interpolating across this model.

estimate the sublimation radius r as:

$$r = \left( \frac{L}{16\pi\sigma T_{sub}^4} \right)^{\frac{1}{2}} \quad (4)$$

where $\sigma$ is the Stephan-Boltzman constant.

Kobayashi et al. (2011) estimate that the sublimation temperature of circumstellar disk dust is between 1300K and 1600K. Using equations 3 and 4, we can predict the orbital distance to inner disk edge and its rotation period for a dipper as a function of stellar luminosity (for any given sublimation temperature).

In both figures, the ONC dipper population, as well as the Hedges et al. (2018)'s sample, sit in a region with longer periods (and wider separations) than would be expected for dust sublimation temperature of 1300–1600K. We overlay upper and lower bounds (corresponding to temperatures of 1300K and 500K) which roughly enclose both samples in Figures 18 and 19. This supports previous findings that dippers possess more extended sublimation radii, and may correspond to the presence of additional heating mechanisms in the inner disk. In line with this, both our findings and those of Hedges et al. 2018 support the magnetospheric accretion model for dippers proposed by Bodman et al. 2017. This prediction that dippers feature lower disk edge temperatures than the dust sublimation temperatures directly relates to the prevalence of dippers among low-mass stars.

### 5.4 The Dipper Fraction

Following the above analysis of the Orion Dipper population present in NGTS, it became pertinent to assess the statistical nature of this population relative to other well-studied populations. These populations are NGC 2264 (Cody et al. 2014), Taurus (Roggero et al. 2021), Upper Sco, and Rho Ophiuchus (Ansdell et al. 2016, Hedges et al. 2018).

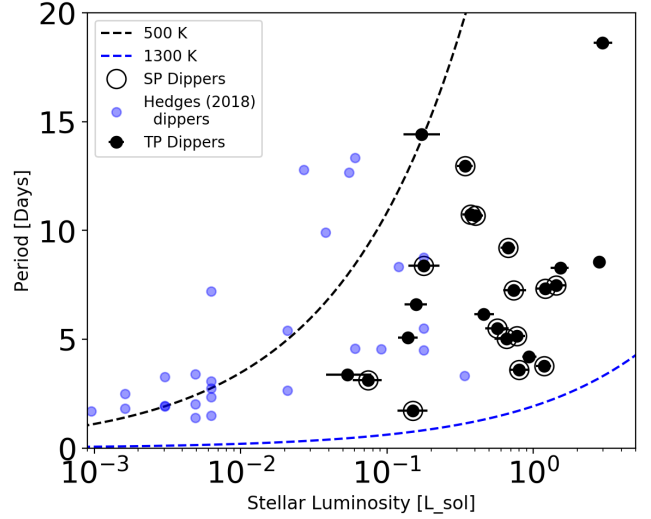

**Figure 18.** Lomb-Scargle period vs stellar luminosity for ONC dippers with both secure and tentative periods. A comparison set of Upper Sco dippers is also show (from Hedges et al. 2018). The dashed lines correspond to expected sublimation radii for dust at 500K and 1300K.

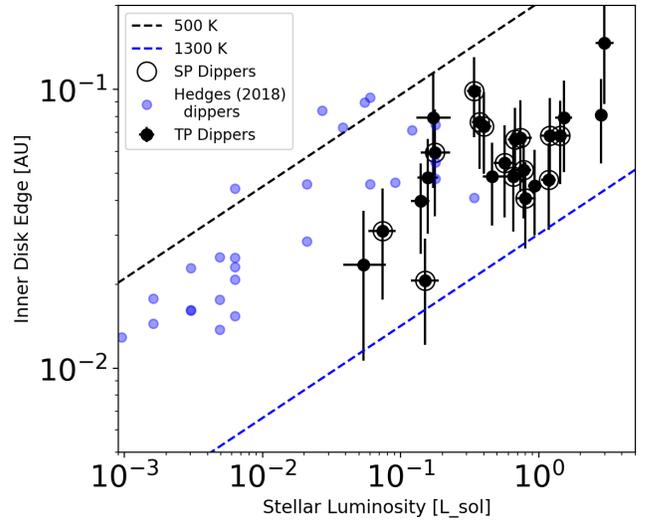

**Figure 19.** Orbital distance to inner disk edge vs stellar luminosity for dippers with both secure and tentative periods. The orbital distance to inner disk edge is calculated following the process outlined in the text, using masses derived from BHAC15. A comparison set of Upper Sco dippers is also show (from Hedges et al. 2018). The dashed lines represent predicted sublimation radii for dust at 500K and 1300K.

One of the highest-order statistics that has been used to characterize cluster-wide Dipper populations is the Dipper Fraction. Defined in Cody et al. (2014) and recycled in Hedges et al. (2018), this metric measures the proportion of disk-bearing stars in a cluster that are dippers. This metric is not only valuable in providing a representation of the dipper occurrence rate, but is also useful in probing the connection between inclination and dipper occurrence hypothesized by McGinnis et al. (2015) and Bodman et al. (2017).

Table 4 presents relevant information on the five well-studied dipper populations, namely the cluster ages, disk fractions, dipper fractions, and dipper number. The Orion dipper population is the largest one observed to date, with the dipper number of 92 greatly exceeding that of the other clusters. The Orion Dipper Fraction is found to be $27.8 \pm 2.9$ per cent. This is in line with the 20-30 per cent occurrence rate predicted by the inclination effect hypothesis, and strongly matches the dipper fraction reflected in the other 4 clusters.

**Table 4.** Dipper fractions for the five regions with well-studied dipper populations, expanded from Hedges et al. (2018) Table 9. We have made the following modifications to this table. (1) We have added error bars to the age estimates based on available literature and the age ranges formerly expressed in Hedges et al. (2018). (2) We have adjusted the disc-bearing fraction for NGC 2264 to reflect more recent findings. (3) Orion has been added in this work, and calculations are further made for young (1-3 Myr) and old (3-10 Myr) members. We further differentiate confirmed Orion members (K18) from the larger population of stars in the Orion NGTS field. The dipper fractions and numbers listed here are a function of disc-bearing stars only. We note the low alignment between the metrics in the final row as stars in the NGTS Orion field are not preselected. References herein are as follows:[1]Luhman & Rieke (1999) [2]Hedges et al. (2002) [4]Venuti et al. (2018) [5]Cody et al. (2014) [6]Pecaut et al. (2012) [7]Kraus & Hillenbrand (2009) [8]Roggero et al. (2021) [9]Megeath et al. (2012) [10]Kounkel et al. (2018)

| Region | Age (Myr) | Disk Fraction (Per cent) | Dipper Fraction (Per cent) | Dipper Number |
|---|---|---|---|---|
| $\rho$ Ophiuchus | $0.5 \pm 0.5$[1] | $40.2 \pm 4.3$[2] | $20.1 \pm 4.3$[2] | 22 |
| Taurus | $1.5 \pm 0.5$[7] | $61.3 \pm 5.8$[8] | $30.9 \pm 5.3$[8] | 34 |
| NGC 2264 | $3 \pm 2$[3] | $28 \pm 2.1$ [4] | $21.6 \pm 3.7$ [5] | 35 |
| Upper Scorpius | $10 \pm 3$[6] | $26.7 \pm 2.0$[2] | $21.8 \pm 3.4$ [2] | 42 |
| NGTS Orion | $2 \pm 1$[9] | $25.8 \pm 1.4$ | $27.8 \pm 2.9$ | 92 |
| NGTS: Young Orion | 0-3[10] | $33.9 \pm 3.4$ | $30.4 \pm 5.5$ | 31 |
| NGTS: Old Orion | 3 - 10[10] | $23.3 \pm 1.5$ | $26.6 \pm 3.4$ | 61 |
| NGTS Orion Field | - | $7.5 \pm 0.3$ | $20.9 \pm 1.8$ | 129 |

### 5.5 Disk and dipper fraction as a function of age

K18 identified observed stars as being members, non-members, or having uncertain membership of the ONC. They also identified which sub-region of the ONC the stars belong to. CMD-derived ages were provided for each sub-region, whereas HR-derived ages were less complete; we used the CMD-derived ages for all member stars.

Previous studies have not yet examined the relationship between dipper fraction and age, likely due to the small number of clusters bearing well-studied populations. In Hedges et al. 2018, the properties of dipper populations in only three clusters were compared. As the Orion dipper population identified in this survey is nearly double that of all previous studied clusters, we used the ages from K18 to split our sample into two groups. These are Young Orion: stars in groups aged less than 3 Myr and largely associated with the Orion A cloud, and Old Orion: stars associated with less active-star forming regions in the complex older than 3 Myr. The resulting disk fractions, dipper numbers, and dipper fractions, can be found in Table 4. Measuring the fraction of stars with disks is of course dependent on the strategy and sensitivity of the survey. All of our sample have based their disk fractions (at least in part) on WISE photometry and so should be comparable, apart from Roggero et al. 2021 which shows a rather higher disk fraction for Taurus than for the other SFRs.

With this division and the addition of Roggero et al. (2021)'s dipper analysis in Taurus, the number of defined stellar regions with well-studied populations has been doubled to 6 when compared to Hedges et al. 2018.

In Fig. 20 we show the disk fraction and dipper fractions, respectively, for the 4 clusters with previously studied dipper populations as well as for Young and Old Orion. In the top panel of Fig. 20, there is a decline in disk fraction with age (though not so steep as published by Haisch et al. 2001, perhaps due to differing methods used in detecting disks). Upper Sco maintains a significant disk fraction at $26.7 \pm 2.0$ for its age of 10 Myr, and lower than the disk fractions observed in $\rho$ Oph and Taurus.

Examining the relationship between age and dipper fraction, it appears from the bottom panel of Fig. 20 that the dipper fraction is independent of age. This finding suggests that at the current age limit of well-studied dipper populations, which is Upper Sco at 10 Myr, the dipper fraction is not found to decline below the inclination effect hypothesis range of 20–30 per cent dipper occurrence McGinnis et al. (2015). This provides an indication that the evolutionary phenomenon of dipping can be frequently observed in clusters up to the age of at least 10 Myr. Our understanding of the relationship between stellar age and the dipping phenomenon would benefit from surveys of older clusters in order to identify whether smaller dipper fractions can possibly be observed, or whether the inclination effect hypothesis holds up until a point in time where such stellar accretion events become negligible within a cluster. This latter possibility may seem unlikely given the age-independent nature of the hypothesis, but cannot be ruled out with current limited information present for dippers.

## 6 SUMMARY

We have used a long baseline photometric survey with NGTS to search for dippers in the Orion Nebula Cluster. The milli-magnitude precision of the lightcurves enabled us to robustly identify 2105 variables with >= 1 per cent amplitude (out of a total sample of 8247 targets).

We employed an iterative Random Forest machine learning algorithm to classify the variable star lightcurves into three distinct classes: periodic variables, eclipsing binaries and dippers. Our training set comprised published examples of each class which fell into our survey dataset. We built a set of 14 features, 13 of which were extracted from the lightcurve photometry, and 1 which used infrared photometry (K-W2) to indicate the presence of a disk.

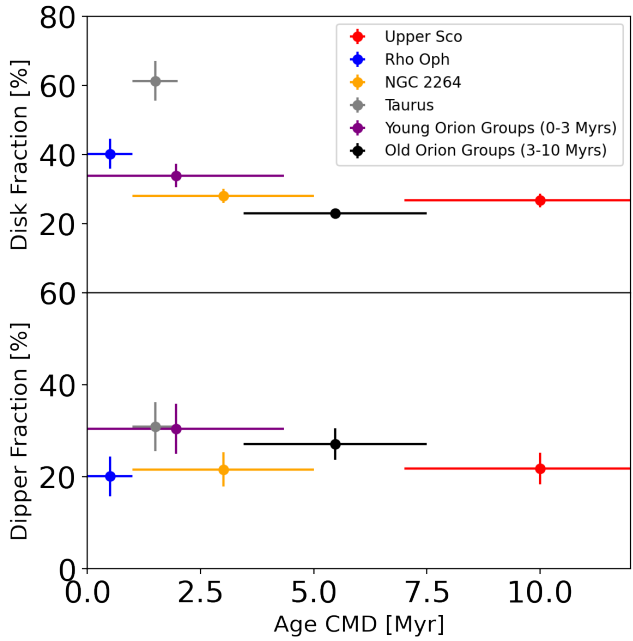

**Figure 20.** Plot of disk fraction (top) and dipper fraction (bottom) as a function of CMD-derived ages for all Orion groups with significant dipper populations. We also show results from the literature for Rho Ophiuchus, Upper Sco, Taurus and NGC 2264. Note that the dipper fraction is given as a fraction of the number of disk-bearing stars. The errors in ages are determined by previous studies for all groups besides Orion; for Orion, these errors are average over the weighted group errors of stars in young and old Orion as determined by K18. Errors on disk fraction and dipper fraction are calculated as simple counting errors.

We found a total of 120 new dippers in the ONC, forming the single largest sample yet discovered in a single star forming region. We also identified 4 new eclipsing binaries, and 1044 periodic variables (with probability>0.9) dominated by rotating spotty stars.

An additional Random Forest algorithm was built and used for the binary classification of objects as either having, or not having, a disk. We applied the algorithm to the K18 sample of ONC stars and found that K-W2 is the single most important feature for disk classification in our sample. Out of the total sample of 129 dippers, 128 were classified by us as disk-bearing, with 1 object being classified as diskless (probably in error). From a combination of near- and mid-infrared photometry, we show that the majority of dippers bear full disks, with a few in transitional–evolved states. With respect to spectral class, the majority of our dippers are found to be K and M type sources, in line with previously discovered dipper populations (e.g. Cody et al. 2014, Hedges et al. 2018). Leveraging findings on the disk and dipper fractions of 4 previously studied clusters and by dividing our Orion dipper sources by age, we are able to compare disk fractions (the fraction of stars bearing disks) and dipper fractions (the fraction of stars with disks with observed dipping events) for 6 populations of dippers distinct in age. We see that despite a declining prevalence of disks in these stars (falling from 40 - 20 per cent), the dipper fraction stays approximately constant. We searched for periodicity in the NGTS lightcurves of dippers using a generalized Lomb-Scargle algorithm. Although the lightcurves of dippers are not well-described by a sinusoid, we found that Lomb-Scargle performed reasonably well for our sample, albeit strongly affected by the window function of the ground-based survey, which results in strong aliasing. Rejection of aliased periods, and eyeballing of phase-folded lightcurves, enabled us to identify that 18 dippers had secure periods, 15 dippers had tentative periods, and 96 had no strong periodic signature. Compared to previous space-based studies of periodic dippers (e.g. Roggero et al. 2021), we find a slightly lower fraction of periodic:non-periodic dippers. This can almost certainly be attributed to the window function of the NGTS survey.

There is a subset of 25 dippers with both spectroscopic measurements of log(g) and Teff, and measured periods (secure and tentative). We used theoretical models to determine masses and luminosities for these stars. Assuming Keplerian orbits we determined distances to the region of the disk where the obscuring material lies, thereby setting constraints on the location of the inner disk edge, and consistent with expected temperatures of around 500–1300K. As seen before (e.g. Roggero et al. 2021), this is significantly cooler than the expected sublimation temperature of dust and adds weight to previous studies that the inner disk edge is somewhat further out than simple model predictions.

## ACKNOWLEDGEMENTS

Based on data collected under the NGTS project at the ESO La Silla Paranal Observatory. The NGTS facility is operated by the consortium institutes with support from the UK Science and Technology Facilities Council (STFC) under projects ST/M001962/1, ST/S002642/1 and ST/W003163/1.

This work was funded by UKRI grants ST/M001962/1, ST/S002642/1 and ST/W003163/1. For the purpose of open access, the authors have applied a creative commons attribution (CC BY) licence to any author accepted manuscript version arising.

This work has made use of data from the European Space Agency (ESA) mission Gaia (https://www.cosmos.esa.int/gaia), processed by the Gaia Data Processing and Analysis Consortium (DPAC, https://www.cosmos.esa.int/web/gaia/dpac/consortium). Funding for the DPAC has been provided by national institutions, in particular the institutions participating in the Gaia Multilateral Agreement. Further details of funding authorities and individuals contributing to the success of the mission is shown at https://gea.esac.esa.int/archive/documentation/GEDR3/Miscellaneous/sec_acknowl/.

This paper made use of the Whole Sky Database (WSDB) created by Sergey Koposov and maintained at the Institute of Astronomy, Cambridge with financial support from the Science and Technology Facilities Council (STFC) and the European Research Council (ERC).

TM gratefully acknowledges support from the Harvard-Cambridge Summer Fellowship, the Harvard Applied Mathematics Department and Professor Alyssa Goodman in the development of this project.

EG gratefully acknowledges support from the David and Claudia Harding Foundation in the form of a Winton Exoplanet Fellowship.

STH is funded by the Science and Technology Facilities Council grant ST/S000623/1

PJW acknowledges support from STFC under consolidated grant ST/T000406/1.

Armagh Observatory & Planetarium is core funded by the Northern Ireland Executive through the Department for Communities.

This research made use of Astropy[1], a community-developed core Python package for Astronomy (Astropy Collaboration et al. 2013, 2018).

---

[1] http://www.astropy.org

Finally, we thank the referee for their thoughtful comments which have significantly improved the paper.

**APPENDIX A: DISK CLASSIFICATION**

Identifying stars as disk-bearing is important when developing a scientifically meaningful discussion of dipper stars. For example, dipping events have been attributed to disk accretion (e.g. Morales-Calderón et al. 2011; Ansdell et al. (2016); Bodman et al. (2017)). In this section we describe a robust method for identifying the disk-bearing nature of the NGTS ONC sample.

For some of these stars, the act of identifying disks has already been completed. Yao et al. (2018) (hereafter Yao18) measured the infrared excess (using the *Spitzer* IRAC 4.5$\mu$m band) of 1431 stars in the Orion A molecular cloud, 724 of which are included in K18. This leaves a significant fraction of the NGTS sample without a measured IR excess.

We used the 724 confirmed (K18) members with Yao18 disk identifications as a training set to build a machine learning classifier for our sample. A large fraction of our stars have near infrared photometry from 2MASS (Skrutskie et al. 2006) as well as WISE W1, W2 and W3 measurements (taken from the ALLWISE catalogue, Cutri et al. 2013). In Fig. A1 we see how K-W3 vs K-W2 alone show great power in separating ONC stars with and without disks.

While an argument might be presented that Fig. A1 is *good enough* for the construction of a simple disk classification criterion, there is a benefit to using a machine learning algorithm to conduct disk classification. Between Gaia, 2MASS, and WISE, we have photometric magnitudes in 10 different pass bands, thus we are more robust against outlier measurements (e.g. blending or variability). There are also a significant number of objects in the overlap region between classes, and indeed scattered into each others` domains. We elected to build a binary Random Forest classifier to detect simply the presence or absence of a disk in our sample stars (see Nguyen et al. (2018) for other approaches using WISE photometry for the identification of debris disk candidates).

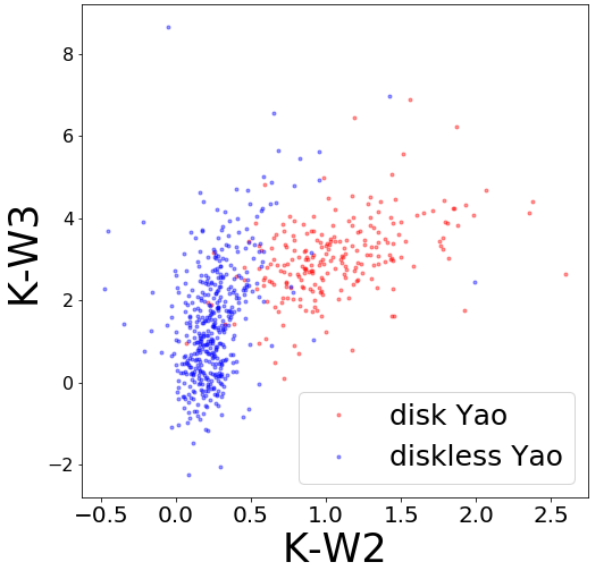

**Figure A1.** Color-Color plot featuring K-W2 vs K-W3 colors for stars in the NGTS Orion dataset previously flagged as diskless and disk-bearing (Yao et al. (2018))

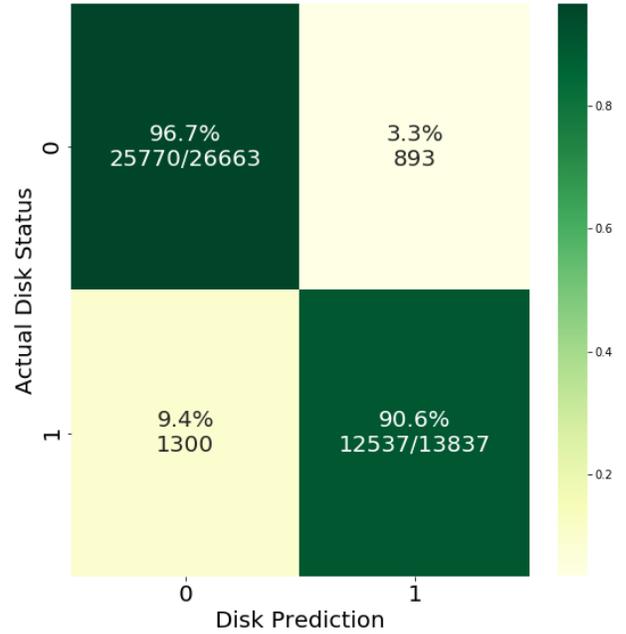

**Figure A2.** Confusion matrix for the disk classifier after 300 iterations using a 80-20 train-test-split.

### A0.1 Disk-Bearing Classification - Building the Training Set and Feature Generation

Our training set is the 724 K18 ONC members with disk status from Yao18 (their labels are 0 for diskless and 1 for disk-bearing). The features we use for each star are simply the baseline Gaia (eDR3) photometry plus a set of colour indexes: $G − BP$, $G − RP$, $G − J$, $G − H$, $G − K$, $K − W1$, $K − W2$, $K − W3$, $K − W4$. Thus our training set consisted of 724 labelled objects with ten features each. Of the 724 stars in our training set, 248 are labelled as disk-bearing.

### A0.2 Disk-Bearing Classification - Building the RF Model

The SKLEARN RF classifier trained in this algorithm used identical parameter tuning for random state, n estimators, and max features as was applied in the morphology classifier (see Section 4). Unlike the variable classifier, very little additional tuning and training was needed. Using a 300-iteration and averaging procedure we obtained the confusion matrix shown in Fig. A2. The model's high recovery rate (over 90 per cent) for both disk bearing and diskless stars indicates the algorithm is quite robust in identifying disk status.

Notably, analyzing the feature importance output of our model in Fig. A3, it is obvious that the infrared excesses – especially the $K − W2$ metric by far – are most important in differentiating stars with and without disks. This validates our initial perception that the infrared colour-colour spaces provide a strong visual illustration of how to isolate stars based on disk-bearing status.

We applied the RF classifier to the remaining 1340 K18 sources. We ran the classifier on each star's set of 10 magnitude and colour features 300 times, and averaged the results. Each iteration produced a set of two probability scores corresponding to the likelihood that the source is either disk-bearing or diskless. We set a simple criterion for the identification of disk-bearing objects – an average disk-bearing probability greater than 0.5. The final set of classified sources is shown in Fig. A4. As we only constructed the classifier to corrobo-rate the disk status of our dipper objects, we have omitted the disk probability scores from our final, online table, although we have included all the data necessary to recreate our results.

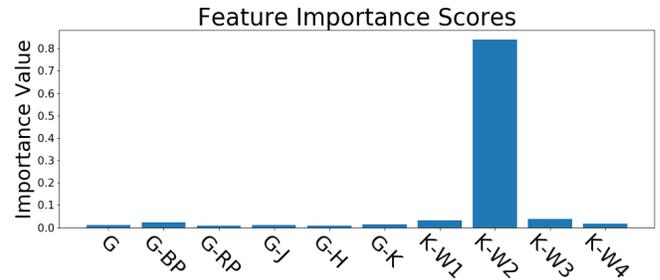

**Figure A3.** Relative importance assigned to each feature in the disk identification RF classifier.

## APPENDIX B: NGTS DIPPER LIGHT CURVES

Figure B1 presents the light curves for the original dipper training set objects used in this study. We direct the reader to the online edition of the paper for the light curves for all 129 dippers used/discovered in this study (Figure B2 and continuing).

This paper has been typeset from a TeX/LaTeX file prepared by the author.

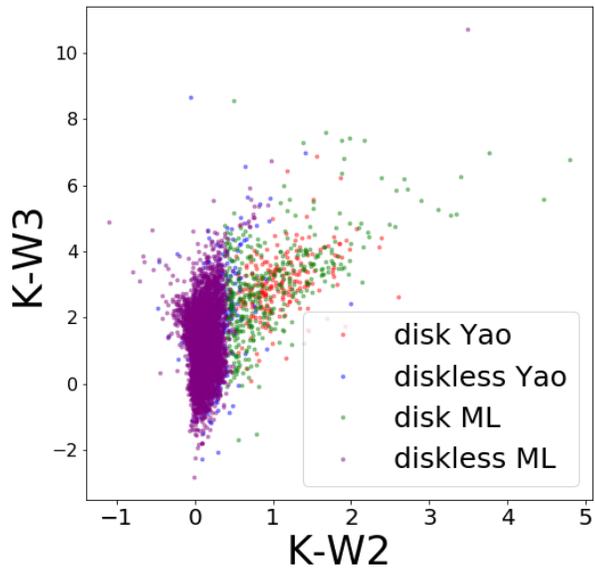

**Figure A4.** Color-Color plot featuring K-W2 vs K-W3 colors for stars in the NGTS Orion dataset previously flagged as diskless and disk-bearing (Yao et al. (2018)) as well as the remaining stars in NGTS as identified by the disk RF classifier.

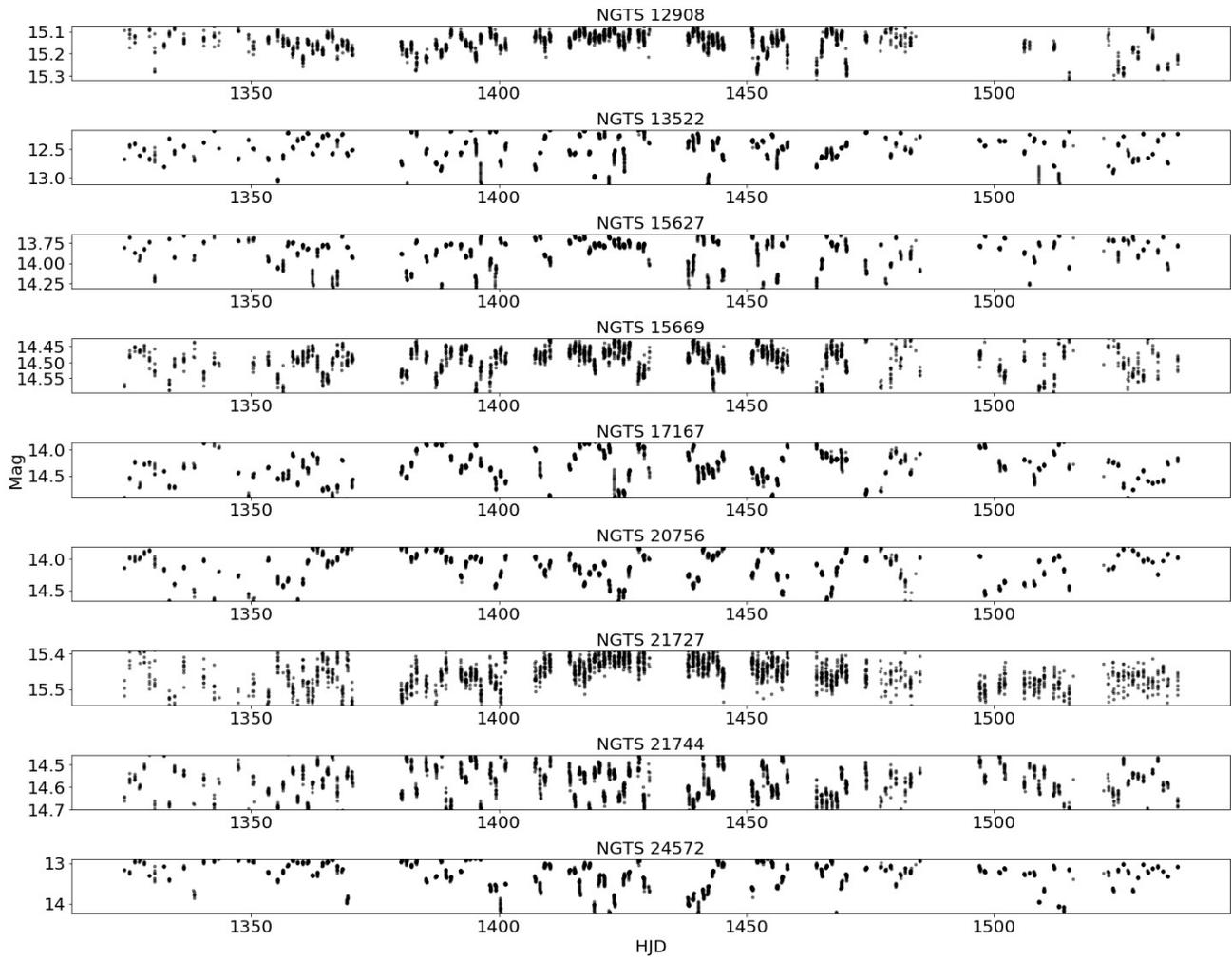

**Figure B1.** Full NGTS light curves for the 9 original training set dippers.

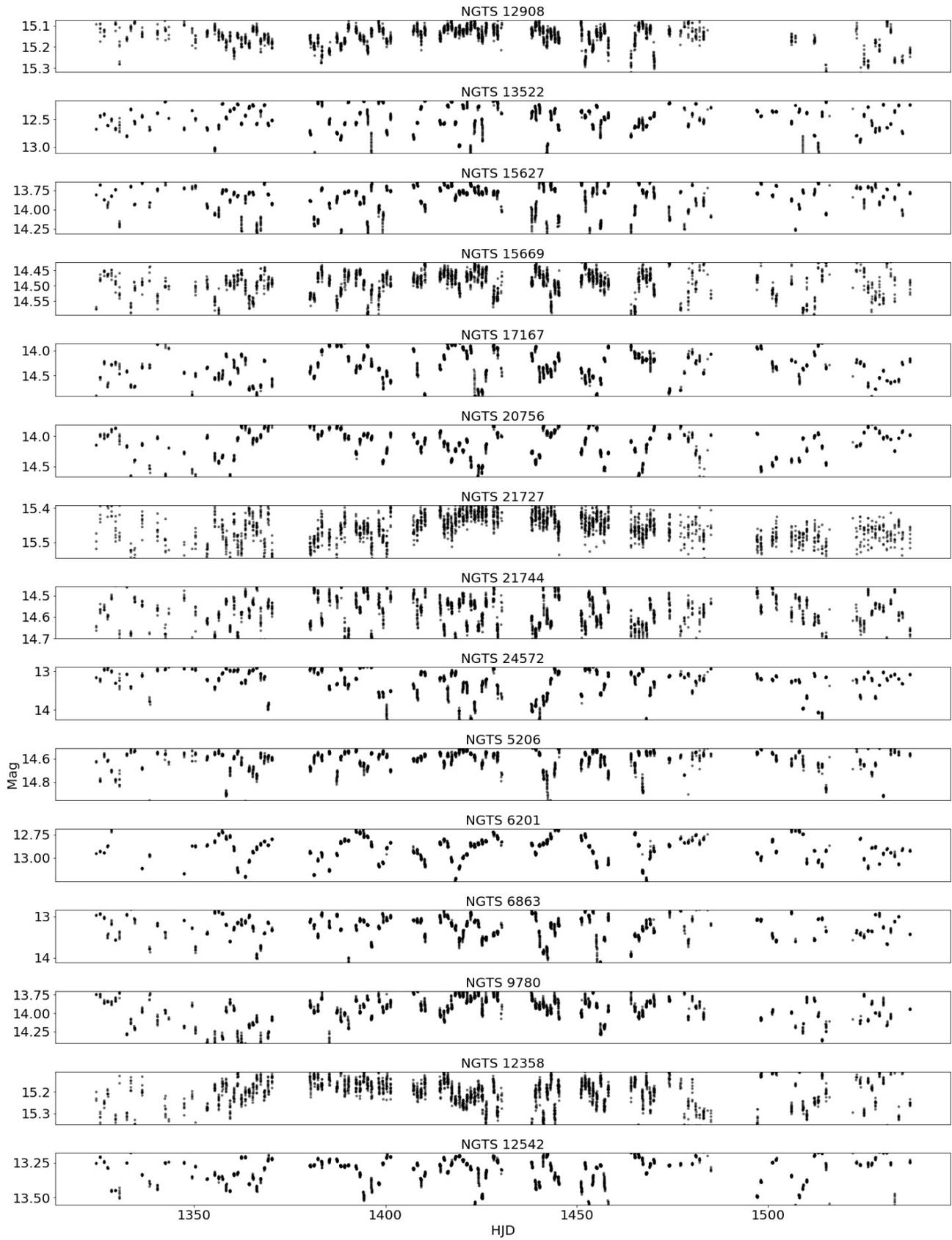

**Figure B2.** Full NGTS light curves for the remaining dippers discovered in the NGTS survey by the authors.

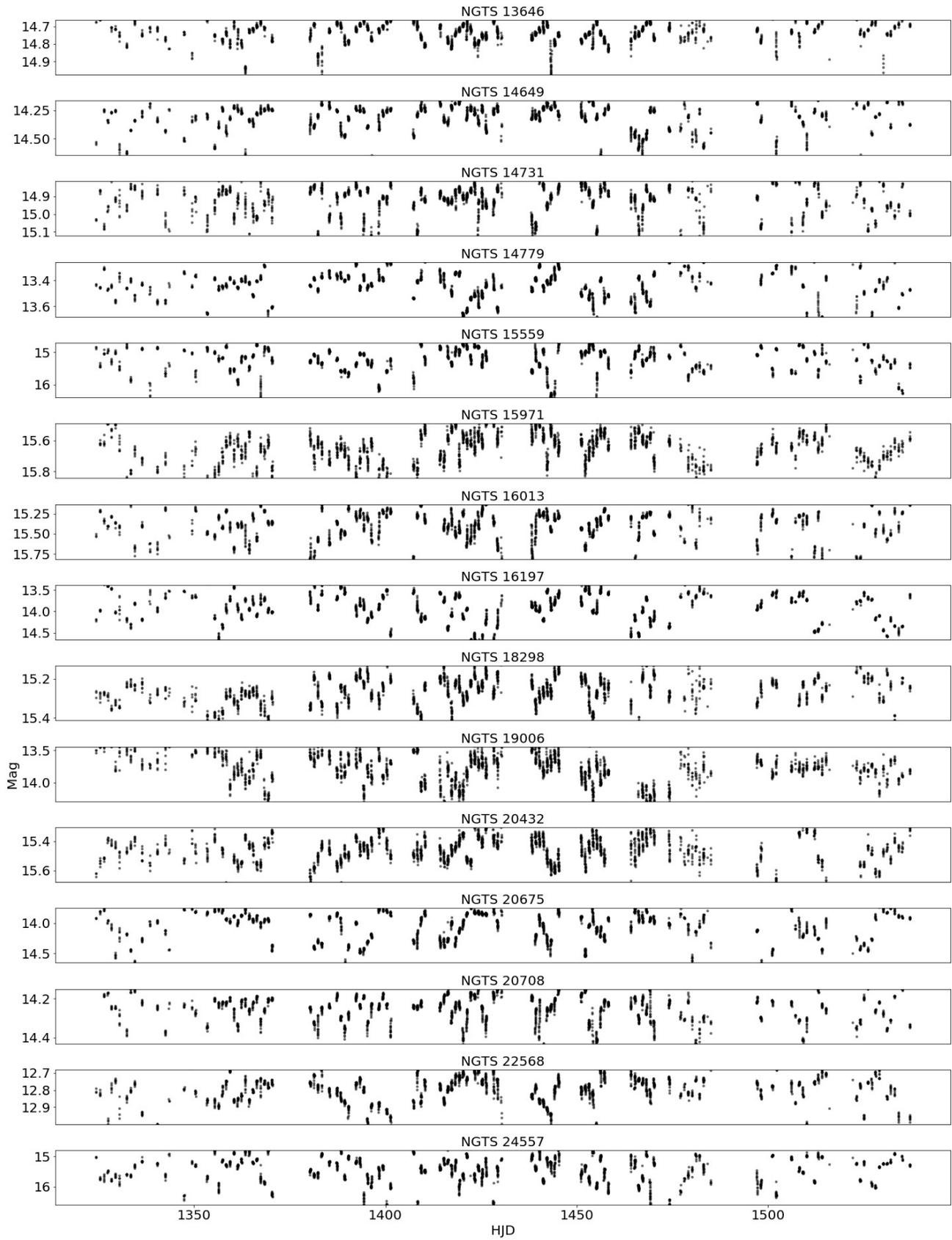

**Figure B3.** Full NGTS light curves for the remaining dippers discovered in the NGTS survey by the authors, continued.

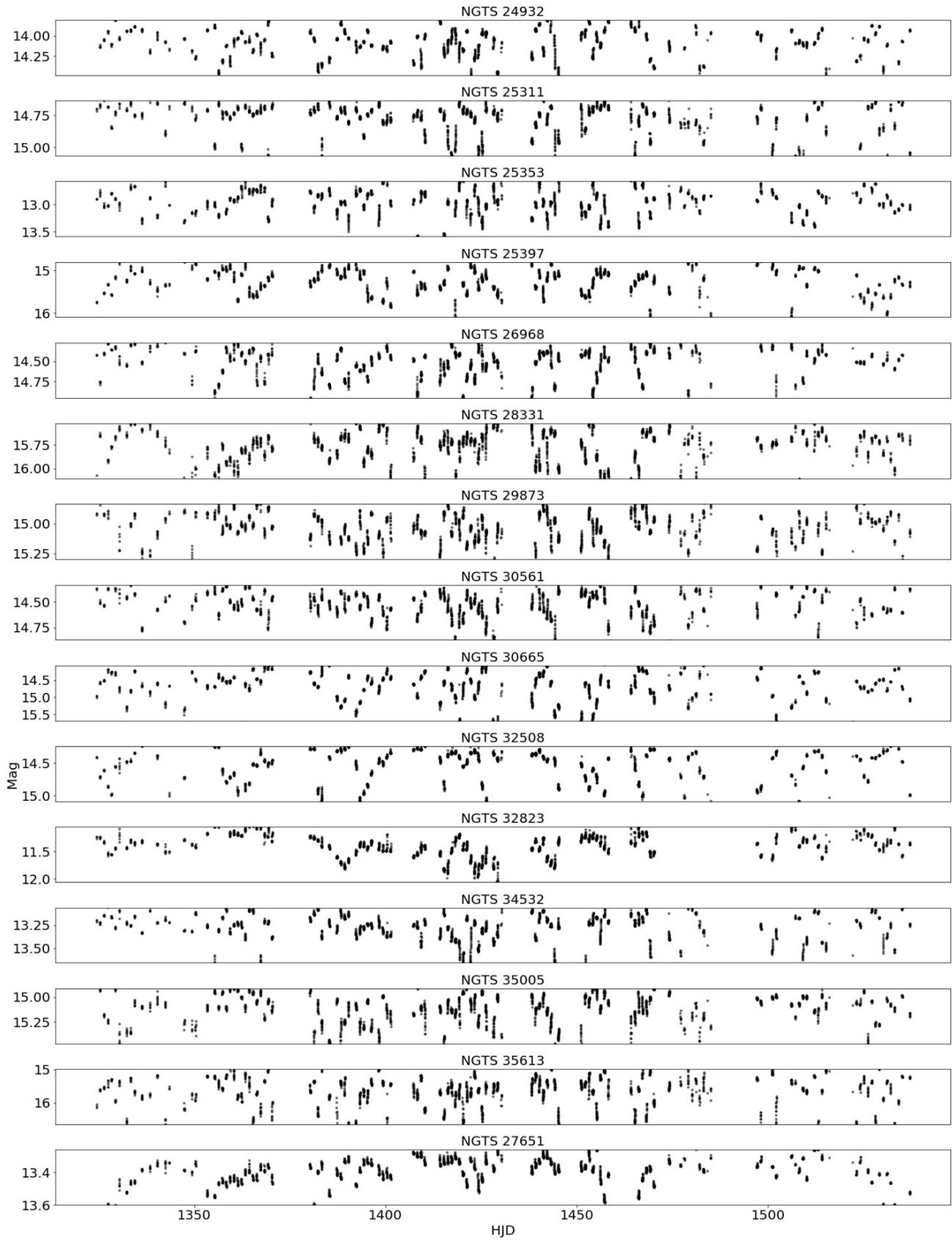

**Figure B4.** Full NGTS light curves for the remaining dippers discovered in the NGTS survey by the authors, continued.

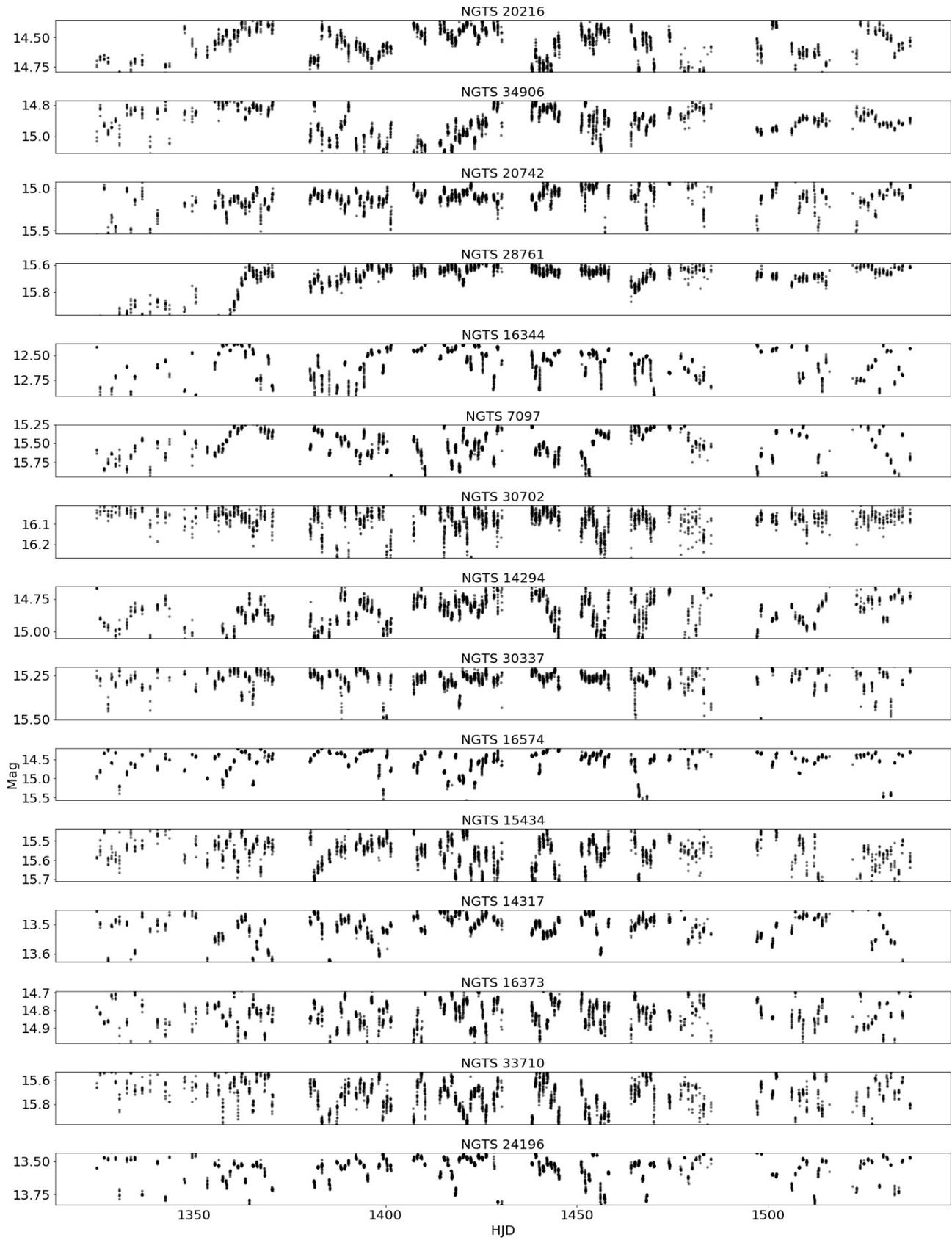

**Figure B5.** Full NGTS light curves for the remaining dippers discovered in the NGTS survey by the authors, continued.

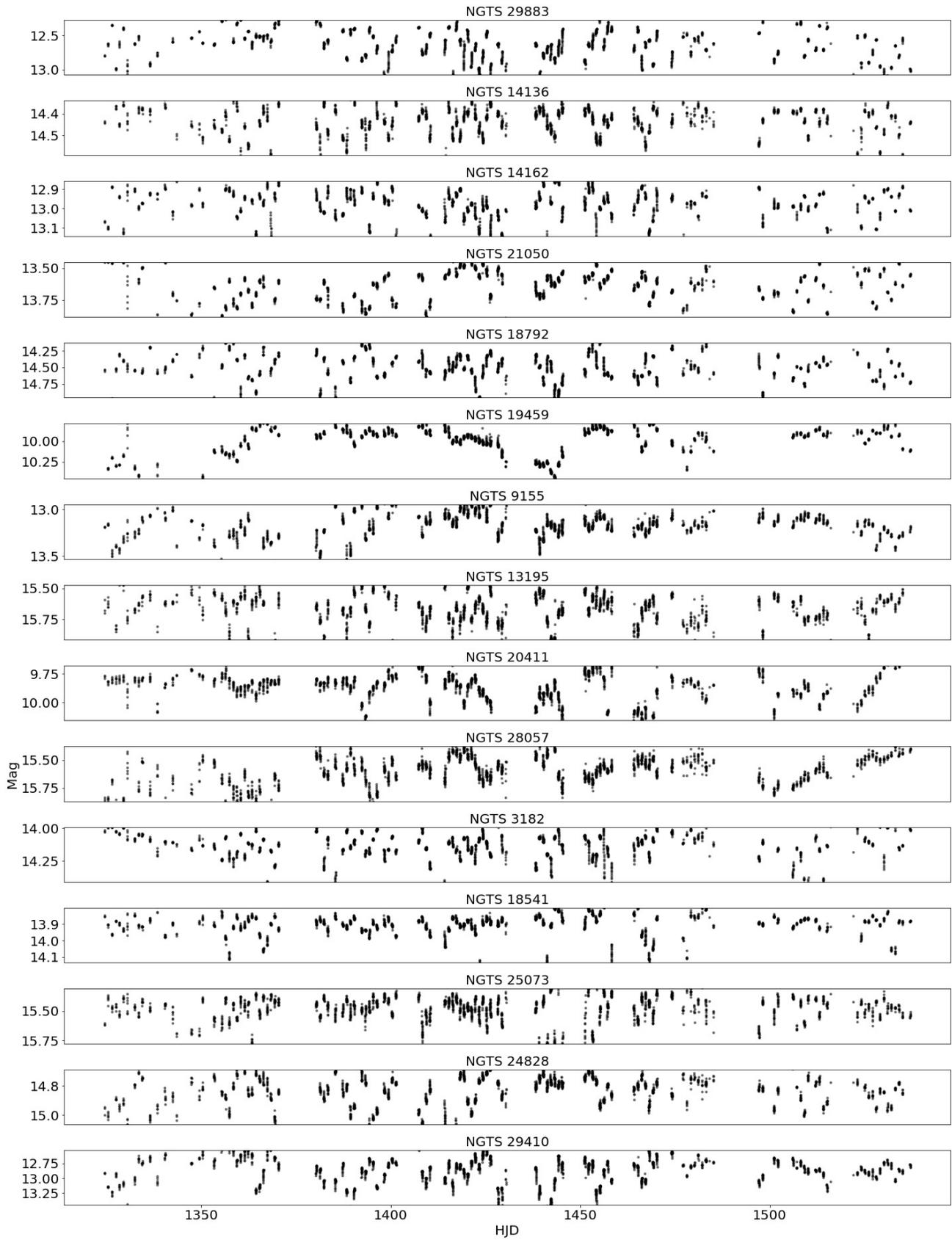

**Figure B6.** Full NGTS light curves for the remaining dippers discovered in the NGTS survey by the authors, continued.

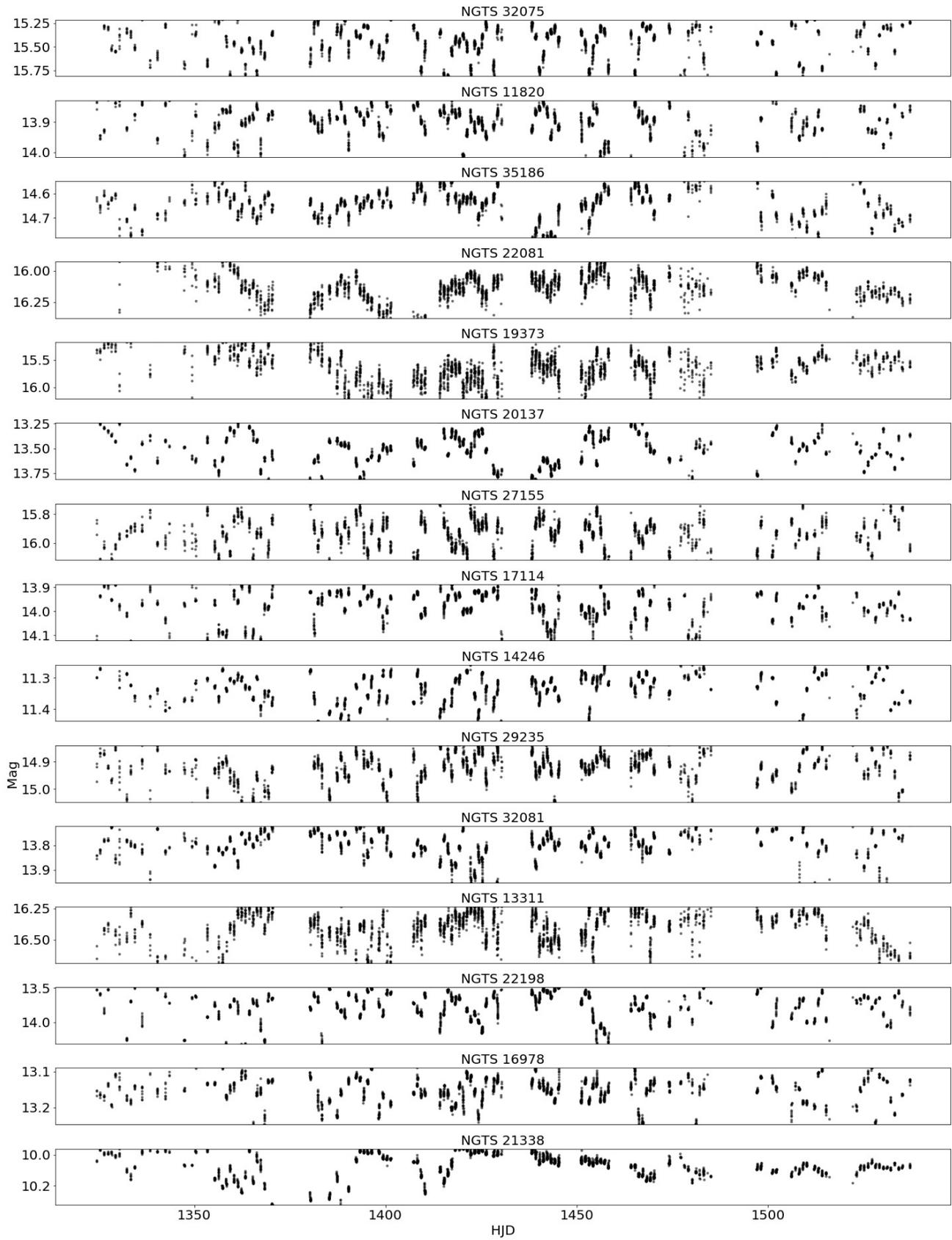

**Figure B7.** Full NGTS light curves for the remaining dippers discovered in the NGTS survey by the authors, continued.

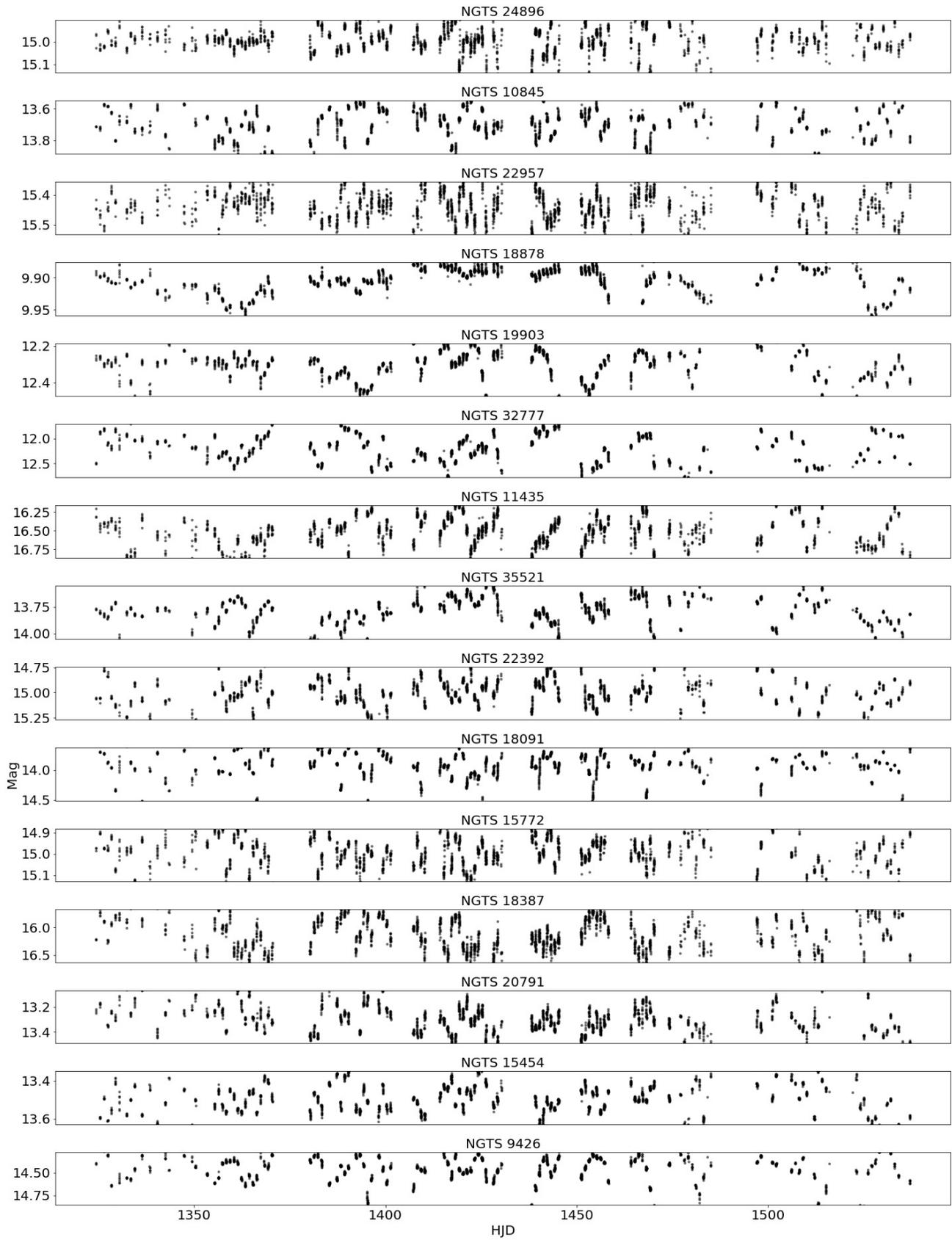

**Figure B8.** Full NGTS light curves for the remaining dippers discovered in the NGTS survey by the authors, continued.

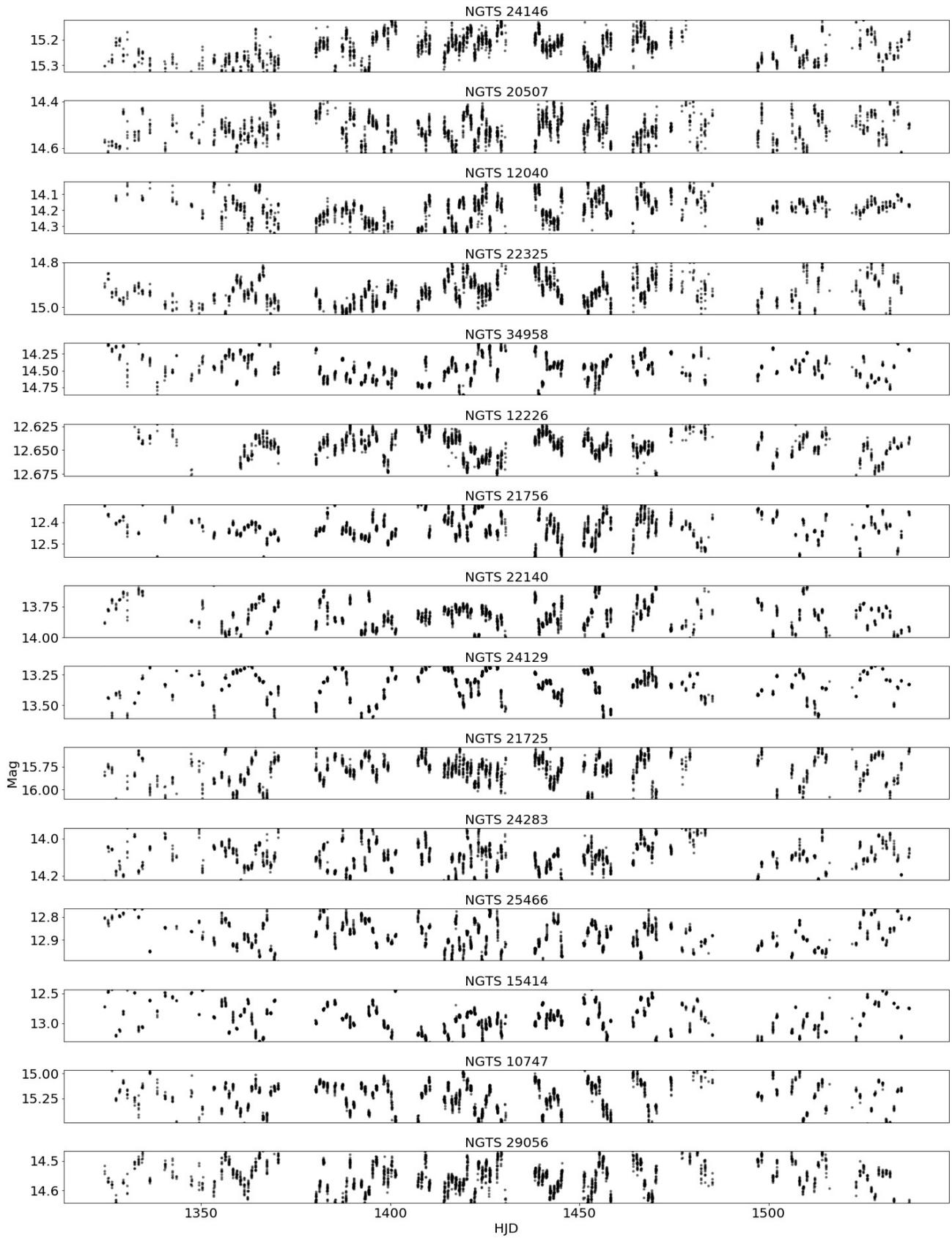

**Figure B9.** Full NGTS light curves for the remaining dippers discovered in the NGTS survey by the authors, continued.

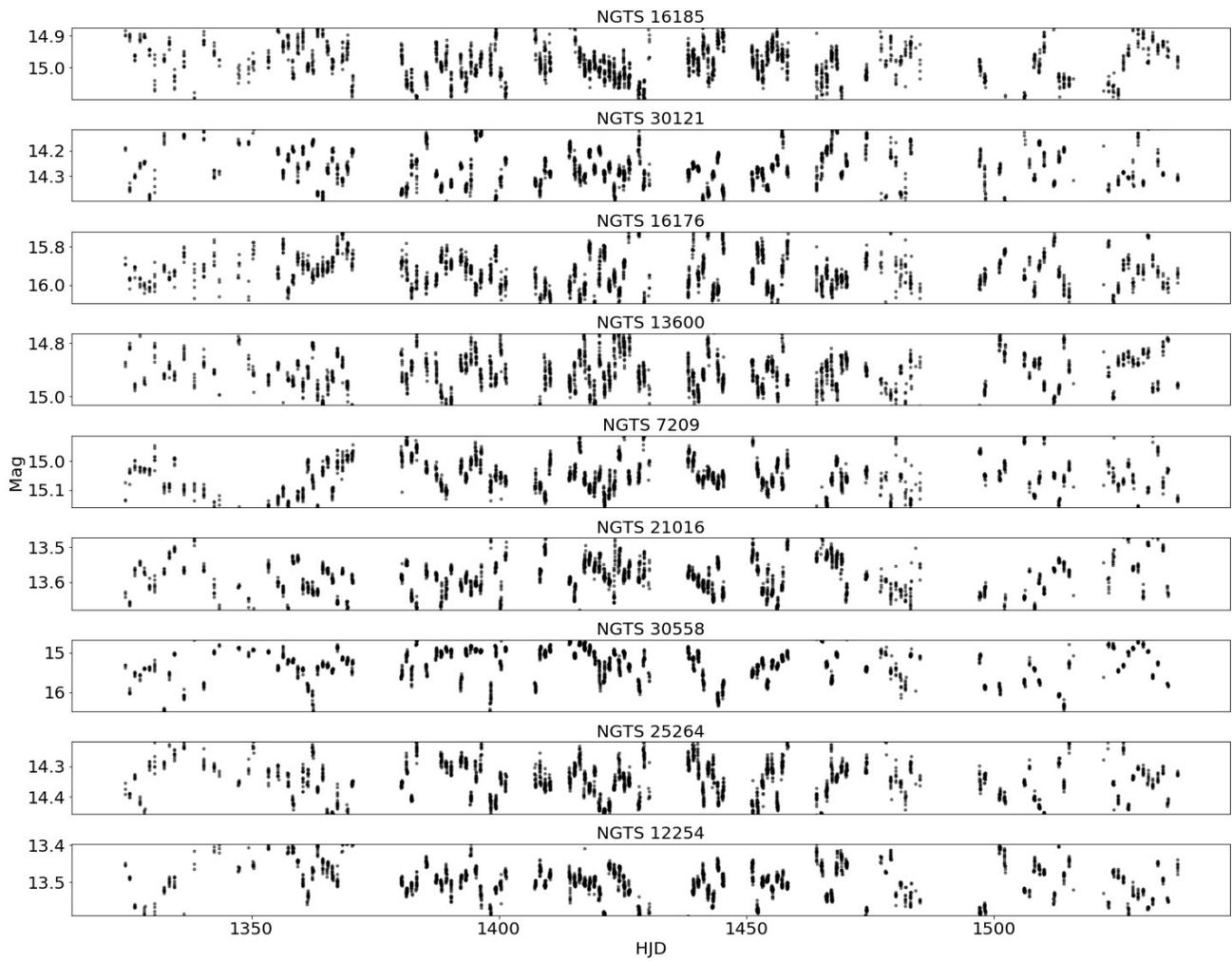

**Figure B10.** Full NGTS light curves for the remaining dippers discovered in the NGTS survey by the authors, continued.